\documentclass[utf8]{frontiersFPHY} 

\setcitestyle{square}
\usepackage{url,hyperref,lineno,microtype,subcaption}
\usepackage[onehalfspacing]{setspace}
\usepackage[nolist,nohyperlinks]{acronym}
\usepackage[dvipsnames]{xcolor}
\usepackage[version=4]{mhchem}
\usepackage{xspace, bm, braket}
\usepackage{siunitx}
\usepackage{mathtools}
\usepackage{dsfont}

\def\keyFont{\fontsize{8}{11}\helveticabold }
\def\firstAuthorLast{H. Chen {et~al.}} 
\def\Authors{Hanghui Chen\,$^{1,2,*}$, Alexander Hampel\,$^{3}$, Jonathan Karp\,$^{4}$, Frank Lechermann\,$^{5}$, and Andrew Millis\,$^{3,6}$}
\def\Address{$^{1}$NYU-ECNU Institute of Physics, NYU Shanghai, 1555 Century Avenue, Shanghai 200122, China \\
$^{2}$Department of Physics, New York University, 726 Broadway, New York, NY 10003, USA \\
$^{3}$Center for Computational Quantum Physics, Flatiron Institute, 162 5th Avenue, New York, NY 10010 \\
$^{4}$ Department of Applied Physics and Applied Mathematics, Columbia University, 538 West 120th Street, New York, NY 10027, USA \\
$^{5}$ European XFEL, Holzkoppel 4, 22869 Schenefeld, Germany \\
$^{6}$ Department of Physics, Columbia University, 538 West 120th Street, New York, NY 10027 USA}

\def\corrAuthor{Hanghui Chen}

\def\corrEmail{hanghui.chen@nyu.edu}

\newif\ifshowcomments
\showcommentstrue

\begin{acronym}[CT-QMC]
 \acro{TMO}{transition metal oxides}
 \acro{CSC}{charge self-consistent}
 \acro{BZ}{Brillouin Zone}
 \acro{CT}{charge transfer}
 \acro{DC}{double counting}
 \acro{DFT}{density functional theory}
 \acro{DMFT}{dynamical mean field theory}
 \acro{FT}{Fourier transform}
 \acro{KS}{Kohn-Sham}
 \acro{MIT}{metal-insulator transition}
 \acro{MLWF}{maximally localized Wannier function}
 \acro{SLWF}{selectively localized Wannier function}
 \acro{OS}{one-shot}
 \acro{QE}{\textsc{Quantum~ESPRESSO}}
 \acro{TB}{tight-binding}
 \acro{VASP}{''Vienna Ab initio Simulation Package''}
 \acro{W90}{\textsc{Wannier90}}
 \acro{WF}{Wannier function}
 \acro{cRPA}{constrained random phase approximation}
 \acro{SIC}{self interaction correction}
 \acro{AIM}{Anderson impurity model}
 \acro{PLO}{projected atomic orbitals}
\end{acronym}

\newcommand{\PNO}{\ce{Pr_4Ni_3O_8}}
\newcommand{\NNO}{\ce{NdNiO_2}}

\newcommand{\ke}{\mathbf{k}}

\begin{document}
\onecolumn
\firstpage{1}

\title[DMFT studies of Infinite Layer Nickelates]{Dynamical Mean Field Studies of Infinite Layer Nickelates: Physics Results and Methodological Implications} 

\author[\firstAuthorLast ]{\Authors} 
\address{} 
\correspondance{} 

\extraAuth{}

\maketitle


\begin{abstract}

\section{}
This article summarizes recent work on the many-body (beyond density functional theory) electronic structure of layered rare-earth nickelates, both in the context of the materials themselves and in comparison to the high-temperature superconducting (high-$T_c$) layered copper-oxide compounds. It aims to outline the current state of our understanding of layered nickelates and to show how the analysis of these fascinating materials can shed light on fundamental questions in modern electronic structure theory. A prime focus is determining how the interacting physics defined over a wide energy range can be estimated and ``downfolded" into a low energy theory that  would describe the relevant degrees of freedom on the $\sim 0.5$\,eV scale and that could be solved to determine superconducting and spin and charge density wave phase boundaries, temperature dependent resistivities, and dynamical susceptibilities. 

\tiny
 \keyFont{ \section{Keywords:} nickelates, correlated electron physics, first-principles, dynamical mean field theory, density functional theory, superconductor} 
\end{abstract}


\section{Introduction}
The identification of a new class of superconductors is a momentous event. Ways in which the new superconductors are similar to or different from previously known materials can drive new insights into the microscopic origin of this  fundamentally mysterious quantum many-body phenomenon.  The discovery ~\cite{Bednorz86,Chu87} of superconductivity in layered copper-oxide materials sparked a revolution in condensed matter physics and materials science, because the transition temperatures were very high relative to other materials known at the time. Additionally, many aspects both of the superconductivity and of the non-superconducting (``normal state'') physics differed sharply from the predictions of conventional theory \cite{Millis00} in ways that made it obvious that interacting electron physics beyond mean field theory could have consequences of fundamental physics interest that approach (and in a few niche cases reach) commercial viability. 

The very recent discovery \cite{Li2019} of superconductivity in the layered $d^9$ nickelates was also momentous because the superconductivity was theoretically anticipated \cite{Anisimov99} on the basis of a chemical and structural similarity to the cuprates. 
Understanding the properties of the superconducting nickelates provides an immense scientific opportunity to sharpen our understanding of the relation between crystal structure and local chemistry on the one hand and important phenomena such as superconductivity on the other. 

The relationship between physical phenomena and crystal structure/local chemistry is the central question in the theory of quantum materials.
The theory of quantum materials is one instance of the quantum many-body problem,  one of the grand scientific challenges of our time. The quantum many-body problem is in essence a problem of  data compression: as is well known,  many-particle quantum mechanics is formulated in a Hilbert space of a size that grows exponentially with the number of degrees of freedom while quantum entanglement in general and Fermi statistics in particular implies that delicate phase relationships between different states are of crucial importance.   Incorporating the chemical realism needed to understand and predict material properties requires acting on this Hilbert space with  a  Hamiltonian matrix that involves a number of parameters which grows as a very high power of the number of degrees of freedom. All solutions of the quantum many-body problem require both reducing the number of degrees of freedom and the number of interaction parameters  to a manageable level (``downfolding" the physics to a simpler systems, typically  a  ``low energy theory" ) and then handling the ``model system" analysis of the  still exponentially large and severely entangled Hilbert space of the downfolded theory.

Three decades of intense work have led to a rough consensus, generally although not universally accepted, that the low energy theory for the copper-oxide materials is some variant of the two dimensional one band Hubbard model. 
The relation of the model parameters to the precise chemistry and structure is reasonably well understood, and the properties of the simplest instantiations of this model are starting to come into focus \cite{simonscollab:2015,Zheng17,Jiang19}.
The recent discovery of superconductivity in layered $d^9$ nickelates such as NdNiO$_2$ takes these questions to a new level. While many aspects of the crystal structure and chemistry are  similar to those of the cuprates, in the nickelate family of materials multiple bands cross the Fermi surface and more interactions may be relevant. A crucial question is whether these effects are  minor, so that the important physics of the nickelates is similar to that of the cuprates, or whether the low energy physics is richer and more complex in nickelates than in cuprates.

Various interacting models have been proposed for NdNiO$_2$. 
The simplest interacting lattice model is a single Ni $d_{x^2-y^2}$ orbital Hubbard model with an additional electron reservoir to mimic the self-doping effect from other bands that exist near the Fermi level. Kitatani \textit{et al.} used this interacting model to directly estimate the superconducting transition temperatures~\cite{Kitatani2020}. Hepting \textit{et al.}~\cite{Hepting2020} and Been \textit{et al.}~\cite{Been2021} used a two-orbital  model including Ni-$d_{x^2-y^2}$ orbital and a lanthanide element $d_{3z^2-r^2}$-like orbital with an interaction only on the Ni-$d_{x^2-y^2}$ orbital. A number of studies~\cite{Hu2019,Werner2020,Zhang2020b,Kang2021,Wang2020b,Wan2021} focus on a different type of two-orbital models that consists of two Ni-$d$ orbitals. Hu \textit{et al.}~\cite{Hu2019} include Ni-$d_{x^2-y^2}$ and Ni-$d_{xy}$ orbitals, while Zhang \textit{et al.}~\cite{Zhang2020b}, Werner \textit{et al.}~\cite{Werner2020} and Kang \textit{et al.}~\cite{Kang2021} and Wan \textit{et al.}~\cite{Wan2021} include Ni-$d_{x^2-y^2}$ and Ni-$d_{3z^2-r^2}$ orbitals. These two-orbital models not only includes the local interaction on each Ni $d$ orbital, but also takes into account the Hund's coupling between the two Ni $d$ orbitals, relevant if the Ni high-spin $S=1$ $d^8$  state is relevant.  Wu \textit{et al.}~\cite{Wu2020} include Ni-$d_{x^2-y^2}$ orbital, neodymium $d_{3z^2-r^2}$ orbital and neodymium $d_{xy}$ orbital. Local interactions are added on both Ni-$d_{x^2-y^2}$ orbital and neodymium $d$ orbitals. This model is used to calculate the spin susceptibility and to estimate the superconducting transition temperature in the weak-coupling limit. Nomura \textit{et al.}~\cite{Nomura2019} compare two different three-orbital basis: one is identical to that of Ref.~\cite{Wu2020} and the other one is Ni $d_{x^2-y^2}$ orbital, lanthanide $d_{3z^2-r^2}$ orbital and interstitial $s$ orbital. Strength of local interactions on all three orbitals is estimated. The model is used to study the screening effects on the Hubbard $U$ of Ni-$d_{x^2-y^2}$ orbital. A different flavor of three-orbital model is employed by Lechermann~\cite{Lechermann2020a}, which consists of Ni $d_{x^2-y^2}$ and Ni $d_{3z^2-r^2}$ orbitals as well as a self-doping orbital. The model emphasizes the importance of multi-orbital processes in $R$NiO$_2$. Gu \textit{et al.}~\cite{Gu2020} use a four-orbital model: Ni $d_{xy}$ orbital, lanthanide $d_{3z^2-r^2}$ orbital, lanthanide $d_{xy}$ orbital and interstitial $s$ orbital. Local interaction is added on Ni-$d_{x^2-y^2}$. The model is used to study the interplay between hybridization and correlation effects and to calculate the phase diagram. Gao \textit{et al.}~\cite{Gao2021} construct a general four-orbital model $B_{1g}@1a \bigoplus A_{1g}@1b$ which consists of two Ni-$d$ orbitals and two lanthanide $d$ orbitals. The model is used to study the topological property of the Fermi surface. All the above models focus on Ni-$d$ and/or lanthanide $d$ orbitals. A number of studies also explicitly take into account oxygen $p$ states with local interactions added on Ni-$d$ orbitals~\cite{Jiang2020,Botana2020,Karp2020,Lechermann2020a}. Jiang \textit{et al.}~\cite{Jiang2020} study a hole doped system that consists of a Ni $d^9$ impurity properly embedded in an infinite square lattice of O $p^6$ ions. This impurity model is used to argue that in the NiO$_2$ layer, the strong $pd$ hybridization may favor a $S=0$ hole-doped state with $^{1}A1$ symmetry, similar to superconducting cuprates. 

Which of the plethora of theoretical models is most relevant is an important question. This review will present a perspective on what has been learned from the \ac{DFT} plus \ac{DMFT} about the many body electronic structure and the relevant low energy model and also what has been learned about the methods in the context of applying them to the layered nickelates.

\section{Overview}

\subsection{Crystal structure and basic chemistry}

\begin{figure}
    \centering
    \includegraphics[width = 0.5 \linewidth]{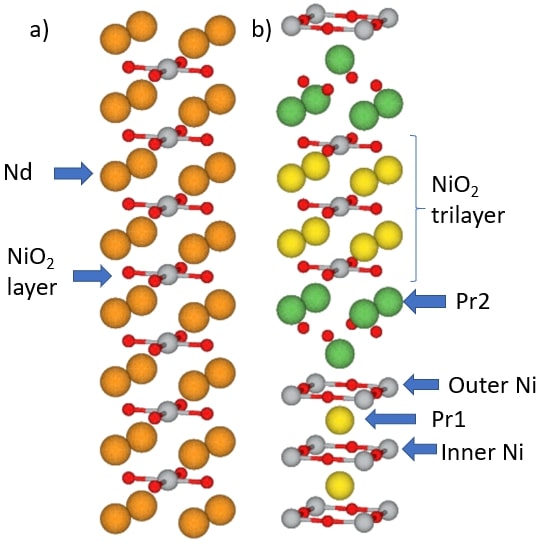}
    \caption{Left: crystal structure of infinite layer \NNO{} in the P4/mmm structure. Right: crystal structure of trilayer \PNO{} in the I4/mmm structure. Ni atoms are shown in silver, O in red, Nd in orange, Pr1 in yellow, and Pr2 in green. From Ref.~\cite{Karp2020} and the crystal structures are visualized using Vesta \cite{vesta}.}
    \label{fig:structs}
\end{figure}

The copper oxide and layered $d^9$ nickelate materials are transition metal oxide (TMO) compounds that share the common structural motif of the CuO$_2$/NiO$_2$ plane.  We focus first on the conceptually simplest materials, ``infinite layer" CaCuO$_2$ and NdNiO$_2$ (panel (a) of Fig.~\ref{fig:structs})  in which the transition metal ions occupy the vertices of a square planar array with the oxygen ions at the midpoints of the edges. In their bulk, three dimensional form these materials  are stacks of transition metal/oxygen planes, with the Ca/Nd ions located half-way between planes, above the centers of the squares formed by 4 transition metal ions. 

The CuO$_2$/NiO$_2$ plane  motif may be combined in many ways, yielding a wide variety of compounds with somewhat similar properties but with many differences of detail. The very large number of cuprate materials will not be discussed here. In the nickelate materials, the infinite layer compounds  can in principle be realized with all rare-earth elements $R$ as A-site cation~\cite{Kapeghian2020}. So far $R=$La, Pr, and Nd have been synthesized. In addition,  multilayer variants $R_{1+m}$Ni$_m O_{2m+2}$ are known,  consisting of groups of $m$ NdNiO$_2$ planes separated by effectively insulating spacer layers of Nd and O. To date, $m=3,4,5$ have been synthesized. The $m=3$ material is shown in panel (b) of Fig~\ref{fig:structs}; the plane labelled Pr2 is the spacer layer.
Some \ac{CT} to the spacer layers occurs, leading to a doping of the NiO$_2$ planes that is different than that of the infinite layer compounds, but apart from this the basic electronic physics of the layered  compounds is believed to be very similar to that of the infinite layer compounds \cite{Karp2020a}. In this article we consider the compounds as interchangeable.

Qualitative chemical (``formal valence") considerations suggest that the electronic configurations of the ions are Ca$^{2+}$Cu$^{2+}\left[{\rm O}^{2-}\right]_2$ and Nd$^{3+}$Ni$^{1+}\left[{\rm O}^{2-}\right]_2$ corresponding in both cases to a transition metal $d^9$ configuration (one hole in the d-shell), and standard considerations of ligand field theory indicate that the hole resides in the transition metal $d_{x^2-y^2}$ orbital. Varying the chemical formula (e.g. by replacing a fraction $x$ of the Nd$^{3+}$ with Sr$^{2+}$ can lead to changes in the Cu/Ni formal valence to $d^{9\pm x}$ (electron or hole doping) and in both material families electron and hole doping leads to dramatic changes in physical properties; in particular, superconductivity occurs only in  relatively narrow electron or hole doping ranges not including the nominal $d^9$ valence.

Formal valence considerations are only a rough approximation to the actual electronic states because in a solid the charge on an ion is not conserved. Quantitatively or qualitatively important roles may be played by charge transfer processes including fluctuations that move an electron from an O to a transition metal (producing a $d^{10} \underbar{L}$ configuration, where $\underbar{L}$ refers to a hole in the oxygen ligand) or from the transition metal to the rare earth/alkali producing e.g. an Ni $d^8$ Nd $^{2+}$ configuration or between two transition metal ions, producing a $d^8-d^{10}$ pair. Especially for Ni based \ac{TMO} it is well known that Ni-O bonds are highly covalent and hybridization effects lead to drastic deviations from the formal atomic orbital occupation picture~\cite{Park12,Mizokawa00}.
The low energy theory of the copper-oxide materials is generally although not universally, accepted to involve one band, of mixed Cu-$d_{x^2-y^2}$ and O$(2p)$ character and with an essentially two dimensional dispersion  and with relatively strong, relatively local interactions derived from intra-$d$ Coulomb matrix elements but substantially affected in magnitude by \ac{CT} and screening processes involving the O ions (the rare earth/alkali ions such as the Ca in CaCuO$_2$ are believed to be electronically inert on these scales) \cite{Millis00}. Important to this finding is the observation that the relevant configurations of the Cu$(3d)$ states are d$^9$ and d$^{10}$; this simple multiplet structure strongly constrains the possible interaction terms. The low energy theory is thus a variant of the two dimensional Hubbard model, possibly including longer ranged interactions. 

In the layered $d^9$ nickelates such as NdNiO$_2$ the situation is richer. Multiple bands cross the Fermi surface, while the potential relevance of $d^8$ configurations raises the possibility that multiplet (``Hund's") interactions governing the relative energetics of different configurations of \ac{TMO} d-electrons are relevant to the low energy physics. 
A fundamental question is whether this additional physics plays an important role in the low energy correlation physics such as superconductivity or whether the correlation physics of the nickelates is in essence similar to that of the cuprates.

\subsection{Downfolding and the DFT+DMFT methodology}

In the context of quantum materials the process of defining an appropriately reduced model from a high energy, more chemically realistic, ``all orbital" description is referred to as ``downfolding''. Downfolding starts from the use of a mean field like approximation such as density functional theory to define a single-particle basis (e.g. the Kohn-Sham eigenstates within a broad energy region) that is for all practical purposes complete. Then  a subspace of this set of states is selected and the Hamiltonian is appropriately projected onto the subspace. The required projection is more sophisticated than simply taking the matrix elements of the Hamiltonian between states in the subspace: screening of the retained interactions by processes involving states that are not retained must be incorporated and correlation induced shifts in relative energy splittings of different states (``double counting corrections") must be managed correctly.  Further, since the purpose is to obtain something that can be solved (perhaps approximately), the specifics of downfolding are intertwined with the method of solution of the resulting theory. 

While one important goal of a downfolding process is to obtain a truly low energy theory (defined, say on the $\pm 0.5$ eV interval around the chemical potential that is relevant for analysis of transport and of  low energy instabilities such as superconductivity), in the case of transition metal oxides an intermediate step of downfolding to a reduced model defined in a wider energy range but with truncated  interactions has been found to be very useful. This intermediate step is motivated by the physical/chemical intuition that in transition metal oxides the  most relevant interactions are the site local charging (``U'') and multiplet (``J") interactions that control the energy levels of a partly filled d shell on a transition metal ion and by the success of the \ac{DMFT} approximation in treating the resulting physics. 

The \ac{DFT}+\ac{DMFT} methodology is a specific downfolding method based on  the idea that both the important interactions and the correlation physics they induce are spatially local in a sense defined more precisely below. It is closely related to  the DFT+$U$ method inmpmllemented  in standard DFT codes; indeed the DFT+$U$ method is in effect the hartreee approximation to the DFT+DMFT method. The locality of the interactions greatly reduces the number of relevant interaction parameters (i.e the complexity of the interacting Hamiltonian that must be solved) while the locality of the correlation physics drastically reduces the severity of entanglement effects, enabling a tractable solution of the resulting correlation problem. The method is now a work-horse of correlated electron materials science and the nickelate materials provide an interesting test of both the assumptions that go into the methodology and the results it produces. 

\begin{figure}
    \centering
    \includegraphics[width = 0.75\linewidth]{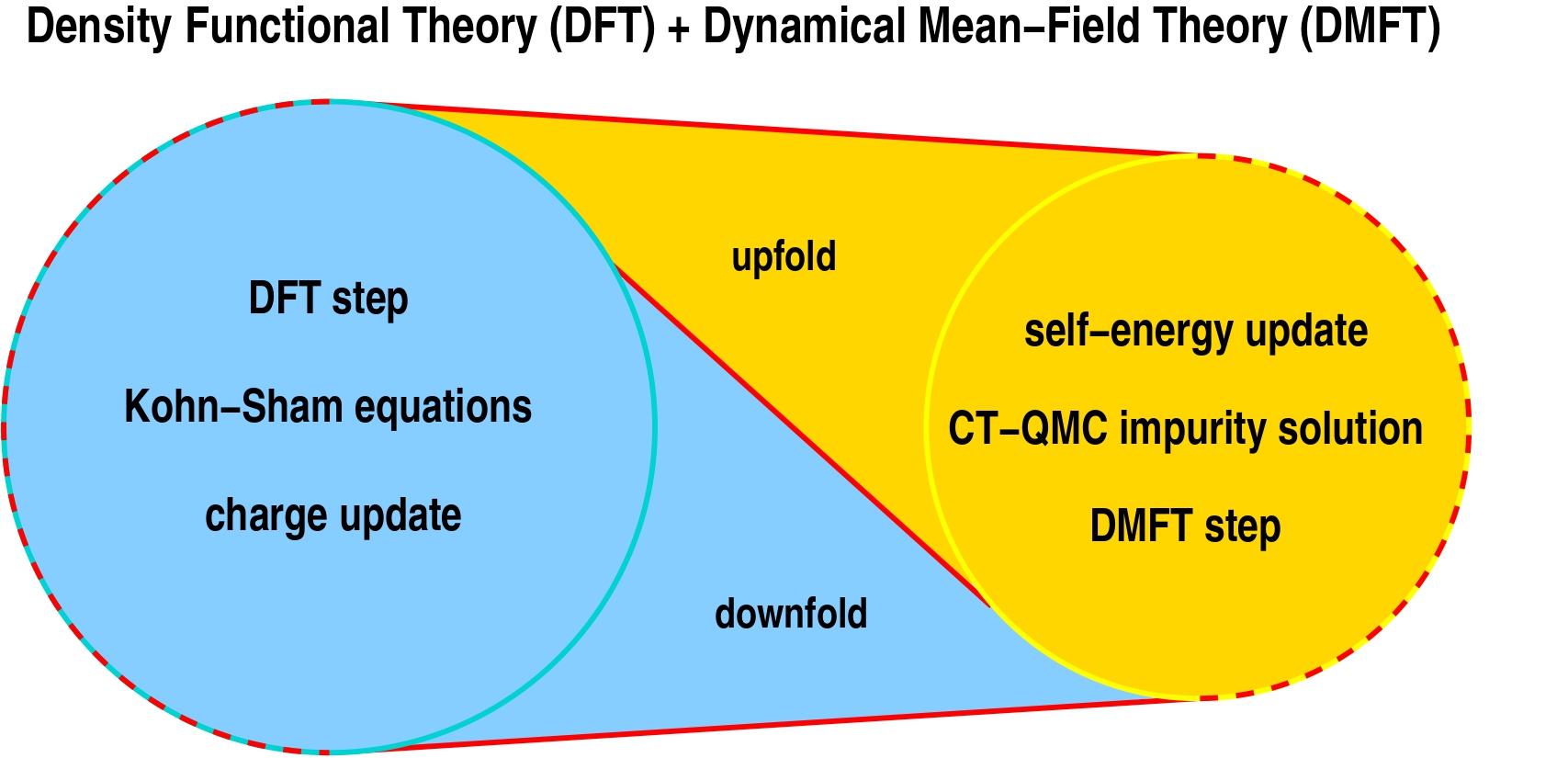}
    \caption{Principle DFT+DMFT method as hybrid scheme, working as a ratchet with up- and downfolding links. One starts from a converged KS-DFT calculation and downfolds to the correlated subspace where a DMFT step is performed. This asks mainly for
    the solution of a quantum-impurity problem (here conducted via
    continuous-time quantum-Monte Carlo (CT-QMC)), yielding a local self-energy. That self-energy is upfolded to the Bloch space of
    the DFT problem and used to revise the electronic charge density. Using the latter, a new KS potential is generated and a novel DFT step performed, which is followed by the next downfolding, etc.. This cycle is performed until self-consistency in the charge density and the self-energy is achieved. Adapted from Ref.~\cite{Lechermann2021a}.}
    \label{fig:dmft_loop}
\end{figure}

The DFT+DMFT methodology ~\cite{Georges04,Kotliar2006} uses a \ac{DFT} calculation to define a set of correlated orbitals (for example the Ni d-orbitals), the hybridization of these orbitals to each other and to other  orbitals, presumed to be uncorrelated, and also involves a prescription for associating site local interactions to the correlated orbitals. 
The resulting downfolded system may be thought of a kind of generalized multi-orbital Hubbard model involving correlated sites coupled to uncorrelated ones and is solved in the \ac{DMFT} (locality of correlation physics) approximation  which reduces the problem to the solution of a quantum impurity model (set of local orbitals coupled to a non-interacting bath) plus a self-consistency condition. 
Finally, in a \ac{CSC} step which is sometimes omitted, the full charge density of the system is obtained and fed back in to the density functional theory; the correlated orbitals are redefined, and the process is repeated until complete self consistency is obtained. This process is visualized in Fig.~\ref{fig:dmft_loop}.
Importantly,the \ac{CSC} \ac{DFT}+\ac{DMFT} equations can be derived as the stationarity conditions of a general functional describing the electronic structure \textit{ab-initio}~\cite{Kotliar2006} within the \ac{DMFT} approximations. 

In first instance, \ac{DMFT} delivers results for the electron (one-particle)  Green's function $\hat{G}$, which is directly connected to the physical spectral function:
\begin{align}
    \hat{A}(\omega) = i\left[\hat{G}(\omega) - \hat{G}(\omega)^\dagger \right]/2\pi \quad .
\end{align}
From solving the impurity problem within \ac{DMFT} the atomic multiplet nature of the correlated manifold can also be analyzed from the many-body density matrix. 
The computed spectral functions can be compared directly to photoemission measurements~\cite{Tamai:2019,Karp:2020} and considerable intuition about the physics can be gained from the many body density matrices.
Furthermore, symmetry broken phases such as superconductivity and magnetic phase can be addressed, and transport coefficients can be estimated. 
Momentum dependent susceptibilities and vertex functions can also be constructed, albeit with much more computational effort. 

The important technical steps in implementing the dynamical mean field approximation are the construction of the correlated orbitals and their coupling to the uncorrelated orbitals and the computation of the interactions among electrons in these orbitals.

\section{Methods}
\label{sec:methods}

The DFT+DMFT methodology is formulated in terms of the  single particle Green function, defined on the Matsubara (imaginary frequency) axis for a periodic solid as
\begin{align}
    \hat G(\ke,i\omega_n)=\left[i\omega_n + \mu - \hat H_{ref}(\ke)- \hat \Sigma(\ke,i\omega_n)\right]^{-1}
    \label{Gdef}
\end{align}
Here $\ke$ is a wavevector in the first Brillouin zone of the solid, $\mu$ is the chemical potential, $H_{ref}$ is a reference single-particle (non-interacting) Hamiltonian and the self energy operator $\hat{\Sigma}$ parametrizes the difference between physical electron propagation and the predictions of the reference single-particle Hamiltonian. If the theory is solved exactly, the choice of reference Hamiltonian is immaterial: it just provides a starting point for calculations. However, actual calculations involve approximations and require the use of a computational basis that is a subset of the full set of all single-particle eigenstates of $H_{ref}$; in this circumstance, the choice of $H_{ref}$ and the choice of basis functions will influence the final result.    

Here, we take $H_{ref}$ to be the Kohn-Sham Hamiltonian produced by a specified density functional theory combined with a ``\ac{DC} correction" and $\hat{\Sigma}^{\nu,\nu^\prime}_\ke$ is a matrix in the space spanned by the eigenstates $\psi_{\nu \ke}(r)$ of $H_{ref}$, which is constructed using a locality ansatz described more explicitly below.
The  different `pure' DFT exchange-correlation functionals give very similar results and will not be discussed explicitly here. 
However, extensions of DFT such as \ac{SIC} DFT~\cite{Perdew1981} do lead to different results, as discussed below. 
For further reading we refer to review articles, i.e. Ref.~\cite{Burke2013}.

In embedding theories such as \ac{DFT}+\ac{DMFT} one defines on physical grounds a subset of correlated orbitals (in the cuprate/nickelate context the transition metal d-orbitals) and retains only the site-local matrix elements of the self energy among these orbitals and the site-local interactions that contribute to these matrix elements.
These site-local self energy matrix elements are then calculated via a mapping to a quantum impurity model with the local Hamiltonian (crystal field levels plus interactions) of the full model and a self-consistently determined coupling to a bath in the form of a so called ``hybridization function''; the result is self-consistently embedded into the full electronic structure.

The key conceptual issues in the method are the construction of the local orbitals and the specification of the interactions between them. These issues are discussed in the next two subsections.

\subsection{Quantum embedding: Construction of the Localized Orbitals}
\label{sec:embedding}

The first step in this procedure is to precisely define a set of orbitals $\phi^{R_i}_m$ centered on sites $R_i$.
These orbitals define a basis in which the self energy and local interactions are computed. 
Within the \ac{DFT}+\ac{DMFT} community many approaches to define these local orbitals have been introduced. 
All methods explicitly or implicitly use a set of so called ``projector functions'':
\begin{equation}\label{eq:projectors}
        P_{\nu,\ke}^{R_i, m}=\braket{\phi^{R_i}_m | \psi_{\nu,\ke}} \quad ,
    \end{equation}
which specify the relation of the local orbitals $\phi^{R_i}_m$ to the Kohn-Sham (KS) eigenstates $\psi_{\nu,\ke}$.
The $P$ operators are termed projectors because in the general case there are fewer localized orbitals $m$ than there are Kohn-Sham eigenstates $\nu$.  

The projector functions allow one to `upfold' a self energy calculated in the orbital ($R_i,m$) basis to the Kohn-Sham basis as
\begin{equation}
    \Sigma_{\nu\nu^\prime}(\ke,i\omega_n)= \sum_{R_i m m'} \left(P_{m \nu }^{R_i \ke}\right)^\star  \left[\Sigma_{\text{QI}}^{R_i} (i\omega_n) \right]_{mm'} P_{\nu^\prime m^\prime}^{R_i \ke} \ .
    \label{sigmaupfold}
    \end{equation}
Here we have written Eq. ~\ref{sigmaupfold} for the DMFT ansatz of a self energy that is site local in the orbital basis. In general the self-energy can also contain multiple sites embedded as a cluster impurity. However, in this case one has to carefully consider breaking of symmetries due to the geometry of the lattice.

The full Green function Eq.~\ref{Gdef}, a matrix in the set of Kohn-Sham bands, is then written as
\begin{equation}
    \left[G^{latt}_{\nu\nu^\prime}(\ke,i\omega_n)\right] =\left[i\omega_n  + \mu  - \hat H_{ref}(\ke)-  \Sigma_{\nu \nu^\prime}(\ke,i\omega_n)\right]^{-1}
    \label{eq:Glatt}
    \end{equation}
The `downfolded' site-local lattice Green function $G^{mm^\prime}_{\text{QI};R_i}$ in the orbital basis is given as
    \begin{eqnarray}
G^{mm^\prime}_{\text{QI};R_i}(i\omega_n)= \frac{1}{N_k} \sum_{\nu,\nu^\prime,\ke \in \mathcal{W}(k)} P_{\nu m}^{R_i \ke } G^{latt}_{\nu,\nu^\prime}(\ke,i\omega_n) \left(P_{m' \nu^\prime }^{R_i \ke}\right)^\star \ ,
\label{eq:SCE1}
\end{eqnarray}
which allows to construct the dynamic Weiss field of the quantum impurity problem in \ac{DMFT} via a Dyson equation: 
\begin{align}
     \mathcal{G}^0_{m m'} (i \omega_n)^{-1} = G^{mm^\prime}_{\text{QI};R_i} (i \omega_n)^{-1} + \Sigma_{\text{QI}}^{R_i} (i \omega_n) \quad .
\label{eq:DMFT_dyson}
\end{align}
$\mathcal{G}^0_{m m'}$ connects the impurity and the bath, from which a hybridization function for the \ac{AIM} can be constructed. 

Solving the impurity problem gives a new self-energy (eq.~\ref{sigmaupfold}), defining a new lattice Green function, and hence a new local Green function. This procedure is performed until convergence is reached and $\mathcal{G}^0$ does not change further.

The \ac{DMFT} converged lattice Green function Eq.~\ref{eq:Glatt} can be used to construct a new charge density:
\begin{equation}
    n(r) = \frac{1}{\beta} \frac{1}{N_k} \sum_{\nu, \ke} \braket{r | \Psi_{\nu,\ke}}  \left[G^{latt}(\ke,i\omega_n)\right]_{\nu\nu} \braket{\Psi_{\nu,\ke} | r} \ ,
\end{equation}
which serves as input for \ac{DFT} in a \ac{CSC} DFT+DMFT calculation.
From this a new charge density from DMFT a new set of \ac{KS} orbitals, projectors and interactions are defined and the \ac{DMFT} equations are solved again. 
This procedure is repeated until the charge density converges. 
The effect of this additional charge self consistency  loop is sometimes small, but in many cases can have important quantitative influence in the results~\cite{Hampel/Beck/Ederer:2020}, and is technically needed to evaluate energies within the DFT+DMFT formalism~\cite{Kotliar2006}.

\subsubsection{Projected atomic orbitals}

One choice for constructing the projector functions is the \ac{PLO} methodology \cite{anisimov2005full,amadon2008plane,aichhorn2009}, in which one introduces a set of atomic-like correlated orbitals $\ket{\tilde \phi_m^{R}}$, which are centered on the positions $R$ of the ions of interest, have the symmetry appropriate to the correlated orbital of interest (e.g. transition metal $d$), and are set to zero for distances $|\vec{r}-\vec{R}|$ around $R$ greater than some preset value. 
In this formalism a first set of projectors is then defined from Eq.~\ref{eq:projectors} as 
\begin{equation}\label{eq:projectors1}
        \tilde{P}_{\nu,\ke}^{R_i, m}=\braket{\tilde{\phi}^{R_i}_m | \psi_{\nu,\ke}} \quad .
    \end{equation}
The downfolded orbitals describing the correlated subspace are computed as
\begin{equation}
        \ket{\bar{\phi}^{R_i}_m} =\sum_{\nu,\ke\in \mathcal{W}(\ke)} \tilde P_{\nu,\ke}^{R_i, m}\ket{\psi_{\nu,\ke}} \ ,
        \label{eq:Pdef1}
\end{equation}  
here, $\mathcal{W}(\ke)$ notates the fact that all practical calculations retain only a subset of \ac{KS} states within a possibly $\ke$-dependent  window $\mathcal{W}(\ke)$. Since the sum over $\nu$ is not complete, as it runs only over $\mathcal{W}(\ke)$, the $\ket{\bar{\phi}^{R_i}_m}$ must be orthonormalized. The result after orthonormalization are functions $\ket{\phi^{R_i}_m}$ that deviate to some degree from the originally defined atomic like states like states $\ket{\tilde{\phi}_m^{R_i}}$, and in particular have tails that extend outside the originally defined radius. These functions may be viewed as Wannier functions as defined as in Eq. ~\ref{eq:wannier}, with the procedure described above corresponding to a prescription for constructing some elements of the U matrices.
These new states are then used in Eq.~\ref{eq:projectors} to construct the projectors $P_{\nu,\ke}^{R_i, m}$, which are actually used in the formalism.

The  window $\mathcal{W}(\ke)$ effectively controls how localized the resulting states are, and one strategy is to formulate the problem in as wide a range as feasible, to make the $\ket{\phi^{R_i}_m}$ very localized.
However, a narrower energy window has the advantages of focusing attention on states that are more directly related to the low energy physics of interest and of providing a theory with lower intrinsic energy scales.

\subsubsection{Maximally localized Wannier functions for downfolding} 

One may also obtain the projector functions via the Wannier construction introduced by \citeauthor{Marzari/Vanderbilt:1997}~\cite{Marzari/Vanderbilt:1997}, in which one defines $N$ spatially localized \acp{WF} as Fourier transforms of linear combinations of $N$ \ac{KS} states via
\begin{align}\label{eq:wannier}
     \ket{w_{\alpha}^{R_i}}  =\frac{V}{(2\pi)^3} \int_{\text{BZ}} d \ke \ e^{-i \ke R_i} \sum_{\nu=1}^{N} U_{\nu\alpha}^{*(\ke)} \ket{\Psi_{\nu,\ke}}\, ,
\end{align}
Here the $U_{\nu\alpha}^{*(\ke)}$ are the components of a family of unitary transformations (one at each $\ke$) and are chosen to optimize specific properties of the $ \ket{w_{\alpha}^{R_i}}$.
The most common choice, referred to as the \ac{MLWF} method, is to choose the $U_{\nu\alpha}^{*(\ke)}$ to minimize the mean square spread of all \acp{WF} in the unit cell~\cite{Marzari/Vanderbilt:1997}. Alternatively, one can construct \acp{SLWF} by performing the spread minimization only for certain \acp{WF}~\cite{Wang14}.

A subset   of the Wannier functions, labelled by $m$, are identified with the correlated orbitals and the Wannier construction in Eq.~\ref{eq:wannier} implies that projectors can be identified as
\begin{align}
    P_{m \nu }^{\ke} \coloneqq U_{\mu\alpha=m}^{*(\ke)}  \quad .
\end{align}
This construction is implemented in the software package \ac{W90}~\cite{Marzari2012}. 

The initial seed for $U_{\mu\alpha}^{(k)}$ are typically obtained by  projections on atomic orbitals similar to the \acp{PLO} above. These are orthonormalized, and then the orthonormalized functions are used as a starting point for the spread minimization. The additional optimization process leads to differences between the MLWF and PLO procedures.

The number of Kohn-Sham bands $N_B$ used in the Wannierization process can be chosen similar to the projector method above via a window $\mathcal{W}(\ke)$ which is ideally set to contain a group of bands separated by energy gaps from all other bands in the solid. 
If $N_B$ is larger than the number of desired Wannier orbitals then a disentanglement procedure is performed, producing a set of $N$ optimized Bloch states  $\ket{\Psi^{(\text{opt})}_{\mu,\ke}}$ labelled by $\mu=1...N$: 
\begin{align}
    \ket{\Psi^{(\text{opt})}_{\mu,\ke}} = \sum_{\nu=1}^{N_B^\ke} U_{\mu\nu}^{\text{dis}(\ke)} \ket{\Psi_{\nu,\ke}}\,.
\end{align}
This ensures ``global smoothness of connection'' and an optimal $k$-point connectivity by minimizing the gauge invariant term of the spread of the resulting \acp{WF}~\cite{Souza/Marzari/Vanderbilt:2001}. Afterwards, the spread-minimization  is performed as usual using the $\Psi^{(\text{opt})}$.
This allows to define the projector functions 
as:
\begin{align}\label{eq:p3}
    P_{\alpha \nu }^{\ke} \coloneqq U_{\mu\alpha}^{*(\ke)} U_{\mu\nu}^{\text{dis}(\ke)} \quad .
\end{align}

It is important to note that while the projection  of the KS Hamiltonian onto a given set of Wannier functions may reproduce the Kohn-Sham bands perfectly, different Wannierization choices may lead to different constructions of the orbitals and to different tight binding parameters. This issue is discussed in more detail in Ref.~\cite{Karp2021dependence} 
\subsubsection{Comparison}

The projector and Wannier constructions of the localized orbitals are conceptually very similar, differing in the specifics of how the correlated orbitals are constructed. the projector method is connected in a intuitively appealing manner to the local orbitals of interest (especially if the projection window $\mathcal{W}$ is set to a wide value), and provides a more convenient construction of double counting correction formulas~\cite{Haule:2015}. In most applications to date the projector method is used only to construct the correlated orbitals  needed in the DMFT procedure. Part of the motivation for this choice is to focus the DMFT treatment on the strongly correlated orbitals not well treated by DFT. The method  avoids the expensive and sometimes unstable spread minimization associated with the Wannier construction. 
 
\begin{figure}
\centering
\includegraphics*[width=0.9\linewidth]{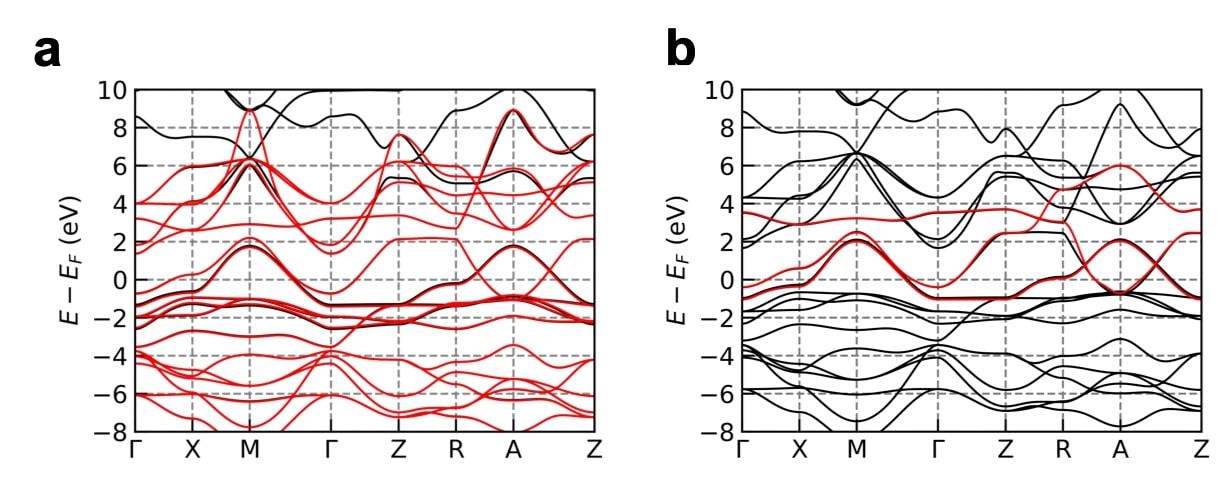}
\caption{\ac{DFT} bands for NdNiO$_2$ (black lines) and Winner fits (red lines) for Wannierization using \textbf{a}  17 Wannier functions (5 Nd-$d$ orbitals + 5 Ni-$d$ orbitals + 6 oxygen-$p$ orbitals + interstitial-$s$ orbital and  \textbf{b} using 3 Wannier functions (Nd-$d_{3z^2-r^2}$, Nd-$d_{xy}$ and Ni-$d_{x^2-y^2}$ orbital). Adapted from Ref.~{\cite{Gu2020}}.}
\label{fig:wannier}
\end{figure}

The \ac{MLWF} method is less dependent on an a-priori choice of atomic orbitals, is based on minimization of a  clear metric under which the orbitals are constructed, and also minimizes deviations from the \ac{KS} dispersion. Because it constructs a basis set which is complete within some energy range, it provides at no additional complication a full tight-binding parametrization of the electronic band structure in a given energy range,
\begin{align}\label{eq:downfolding_hloc}
    H^{\mathrm{TB}}_{\alpha \alpha'}(\ke)= \sum_{\nu} P_{\nu^\prime \alpha'}^{\ke} H_{ref}^{\nu}(\ke) \ \left(P_{\alpha \nu }^{\ke}\right)^\star\ ,
\end{align}  
providing physical intuition about the relevant orbitals. (With some effort, similar information can be obtained from the projector method, but this is not often done). Fig.~\ref{fig:wannier} shows two examples of fitting the NdNiO$_2$ low-energy \ac{KS} states using \ac{MLWF}. The first [Fig.~\ref{fig:wannier}(a)], constructs 17 Wannier orbitals for all 17 \ac{KS} states that are present in a large energy window from -9 to 8 eV, whereas the second [Fig.~\ref{fig:wannier}(b)] constructs a minimal model only for the 3 frontier orbitals (Ni-$d_{x^2-y^2}$, Nd-$d_{z^2}$, Nd-$d_{xy}$). 
As shown in Ref.~\cite{Gu2020} an improved description of the low energy dispersion is obtained by the inclusion of a non-Ni, non-Nd, band orbital near the Fermi level, apparently representing interstitial charge, which is further discussed in Sec.~\ref{subsec:dft_results}. 

In the following we will call all calculations performed directly in the \ac{TB} basis Wannier-mode calculations, whereas calculations in which the \ac{DMFT} equations are written in the Kohn-Sham basis will be referred to as Bloch-mode calculations. Note, that direct formulation of the \ac{DMFT} equations as a solution of a Hamiltonian restricted to the space of  the correlated orbitals is in general not possible, because the projectors are in general non-square matrices and hence do not serve as a unitary transformation~\cite{Beck:2021}.

While Wannier and projector methods had until recently been viewed as roughly equivalent methods of constructing the basis in which the dynamical mean field equations are formulated, and indeed in some cases the equivalence of projector and Wannier-based methods was demonstrated \cite{Park2015},   Karp \textit{et al.} \cite{Karp2021dependence} found that results of the DFT+DMFT methodology can be sensitive to the choice of method used to construct the local orbitals of the downfolded model. We will review and discuss these results in Sec.~\ref{subsec:dmft_local_elec_structure}.

\subsection{Interactions in the correlated subspace} 
\label{sec:interactions}

\subsubsection{Basic definitions}

Once an orbital  downfolding has been defined one has to formulate an appropriate interaction among the downfolded correlated orbitals $\ket{\phi^{R_i}_m}$. Within the DMFT approximation the important interaction matrix elements are the on-site ones among the $n$ correlated orbitals centered on the same site $R_i$.
We begin by considering the matrix elements of the bare Coulomb interaction $V$ among the correlated orbitals on a given site: 
\begin{align}
  \label{eq:uoperator}
  \hat{H}_V = \frac{1}{2} \sum_{\sigma \sigma'} \sum_{m m' m'' m'''} V_{m m' m'' m'''} \ c_{m \sigma}^\dagger c_{m' \sigma'}^\dagger c_{m''' \sigma'} c_{m'' \sigma} \quad .
\end{align}
Here, $c_{m \sigma}^\dagger$ and $c_{m \sigma}$ are creation and annihilation operators for the correlated orbitals $\ket{w_\alpha}=\ket{w^{R_i}_{,m, \sigma }}$ or $\ket{\tilde{\phi}_m^{R_i}}$, and $V_{m m' m'' m'''}$ is the Coulomb interaction tensor:
\begin{align}
  \begin{split}
    \label{eq:u_tensor}
    &V_{m m' m'' m'''} = \bra{m m' } V \ket{m'' m'''} \\
    &= \int d\mathbf{r}_1 \int d\mathbf{r}_2 \ w^*_{m \sigma}(\mathbf{r}_1) w^*_{m' \sigma'}(\mathbf{r}_2) \frac{1}{| \mathbf{r}_1 - \mathbf{r}_2 |} w_{m''' \sigma'}(\mathbf{r}_2) w_{m'' \sigma}(\mathbf{r}_1) \quad .
  \end{split}
\end{align}

In the absence of symmetries there are $\mathcal{O}n^4$ matrix elements, but the main cases of physical interest involve a high degree of symmetry that enables one to reduce the complexity of the Coulomb tensor to just a few parameters. We will only describe the most important formulas here and refer the reader to Ref.~\cite{Pavarini:julich} for an more detailed overview.

The most widely used form is the so called ``Slater'' parametrization~\cite{Slater1960}, which makes use of the spherical symmetry of an isolated single atom. If the projectors in eq.~\ref{eq:projectors} are chosen to produce sufficiently local \acp{WF} this is a good approximation, and is for example used in all DFT+$U$ implementations. 

The most important Coulomb integrals are elements of $V_{m m' m'' m'''}$ that differ only in up to two different indices $m$. Using the symmetries this allows to define (in the absence of spin-orbit coupling):
\begin{align}
  U_{avg} &= \frac{1}{(2l+1)^2} \sum_{m m'} V_{m m' m m'} = F^0  \\
  U_{avg} - J_{avg} &= \frac{1}{2l(2l+1)} \sum_{m \neq m'}  V_{m m' m' m} \quad .
\end{align}
Here, $F^k$, refers to the $k$th Slater integral, proportional in effect to the $k^{th}$-pole of the electric charge distribution of the atomic configuration of the rotationally symmetric free-ion case. From the $F^k$  the full Coulomb tensor can be constructed. For a $d$-shell of an isolated ion one can further show that
\begin{align}
  J_{avg}=\frac{F^2+F^4}{14}
\end{align}
and that $F^4/F^2$ is fixed, so that  the entire Coulomb interaction tensor is determined by only two parameters: $F^0=U_{avg}$ and $J_{avg}$. The ratio $F^4/F^2$ is obtained empirically, varies only little for transition metals, and it is often fixed to $F^4/F^2 \approx 0.625$~\cite{Vaugier2012}. 

Another often applied parametrization is the so called ``Hubbard-Kanamori'' form~\cite{Castellani1978,Kanamori:1963}, widely used in particular to describe the on-site interactions among electrons in the d-shell of a transition metal ion. This parametrization is formulated in terms of the  averaged parameters:
\begin{align}
\begin{split}
  \mathcal{U} &\equiv \frac{1}{n} \sum_{m} V_{mmmm} \\
\mathcal{U}' &\equiv \frac{1}{n(n-1)} \sum_{m \neq m'} V_{mm'mm'} \\
  \mathcal{J} &\equiv \frac{1}{n(n-1)} \sum_{m \neq m'} V_{mm'm'm} \\
  \mathcal{J}_C &\equiv \frac{1}{n(n-1)} \sum_{m \neq m'} V_{mmm'm'} \quad ,
\end{split}
\end{align}
which are the so-called Hubbard-Kanamori parameters for $n$ orbitals. In the particular case of  an octahedral crystal field,  the $d$ shell is split into a $t_{2g}$ and $e_g$ manifold.  Within either the $t_{2g}$ and $e_g$ sub-manifold of the octahedral point group (but not for the whole $d$ shell), one can verify that $\mathcal{U}'=\mathcal{U}-2\mathcal{J}$ and $\mathcal{J}=\mathcal{J}_C$ so as in the free ion case the full  interaction can  be constructed from only two independent parameters $\mathcal{U}$, and $\mathcal{J}$. The resulting interaction operator has the following form:
\begin{align}
\label{eq:Kanamori}
\begin{split}
  \hat{H}_U^{\text{kan}} &= \frac{1}{2} \sum_{\sigma} \sum_{m} \mathcal{U} \ \hat{n}_{m \sigma} \hat{n}_{m \bar{\sigma}} \\
  &+ \frac{1}{2} \sum_{\sigma} \sum_{m \neq m'} \left[ (\mathcal{U}-2\mathcal{J}) \ \hat{n}_{m \sigma} \hat{n}_{m' \bar{\sigma}} + (\mathcal{U}-3\mathcal{J}) \hat{n}_{m \sigma} \hat{n}_{m' \sigma} \right] \\
  &+ \frac{1}{2} \sum_{\sigma} \sum_{m \neq m'} \mathcal{J} ( \ \underbrace{c_{m \sigma}^\dagger c_{m' \bar{\sigma}}^\dagger c_{m \bar{\sigma}} c_{m' \sigma}}_{spin-flip}  + \underbrace{c_{m \sigma}^\dagger c_{m \bar{\sigma}}^\dagger c_{m' \bar{\sigma}} c_{m' \sigma}}_{pair-hopping} ) \ .
\end{split}
\end{align}
Importantly, this form of the interaction is rotationally invariant, which means that within the subset of orbitals, arbitrary unitary transformations on the orbitals can be applied, without adapting the parameters. 

Note, that in this form $\mathcal{U}$ represents directly the diagonal terms of the full Coulomb tensor in contrast to the Slater parameter $F^0$. In the case of spherical symmetry one can show that the  two parametrizations are related by:
\begin{align}
\begin{split}
    \mathcal{U} &= U_{avg} + \frac{8}{7} J_{avg} \\
    \mathcal{J} &= \frac{5}{7} J_{avg} \quad .
\end{split}
\end{align}

\subsubsection{Determining the screened interactions} 
\label{sec:crpa}
In solids the interaction parameters are renormalized from their bare values by screening processes $V \rightarrow U$ involving the other electrons in the solid. The most important renormalization is of the monopole interaction $F_0\equiv U_{avg}$ which gives the charging energy $E_C=\frac{1}{2}U_{avg}N_{tot}^2$ with $N_{tot}=\sum_mn_m$ the total charge in the correlated orbitals. The charging energy measures the change in energy when the local charge is changed, but changing the local charge implies changing the  `monopole' electric fields produced by these charges; these electric fields are screened by the dielectric constant $\epsilon$ so that in computing the energy one should replace $e^2/r$ by $e^2/(\epsilon r)$ in Eq.~\ref{eq:uoperator} where $\epsilon$ is the dielectric constant associated with charge fluctuations on orbitals not included in the low energy theory. 
Since typical values of $\epsilon$ are $\sim 5-10$ renormalizations of the charging energy from the free ion value $\sim 20$ eV to solid state values of the order of $5$ eV are expected. 
The other ``$J$" terms represent energetics associated with electron rearrangement at fixed total charge, i.e. with  changes in the quadrupole and octupole moments of the ion; these fields fall off much more rapidly and the $J$ are as a result much more weakly screened, typically deviating by only $10-20\%$ from the free-ion values. the important conclusion from this argument is that the basic interaction strength $U$ or $F^0$ depends on   how the screening is treated. In the next sections we discuss this issue in more detail.


Commonly used ab-initio methods for treating the strong renormalizations from solid state screening are the constrained LDA method~\cite{Anisimov:1991} and the \ac{cRPA} method~\cite{PhysRevB.70.195104}. Here, we present the latter, for a review see e.g. Ref.~\cite{Aryasetiawan:2011}. 

cRPA is a linear response theory based on the polarization function $P$ giving the response of electrons to a test charge in the system. Within the RPA approximation, which neglects all non-Hartree terms, the dielectric function is calculated from the full bare Coulomb interaction and the polarization function as: 
\begin{align}
  \epsilon = 1 - V P \ .
  \label{eq:full_eps}
\end{align}
The polarization function $P$ in RPA is calculated from DFT as:
\begin{align}
\begin{split}
    P(\mathbf{r},\mathbf{r'},\omega) = \sum_{\nu \mathbf{k}}^{\text{occ}} \sum_{\nu' \mathbf{k}'}^{\text{unocc}} &\bigg[ \frac{ \Psi^\dagger_{\nu \mathbf{k}} (\mathbf{r}) \Psi_{\nu' \mathbf{k}'}(\mathbf{r}) \Psi^\dagger_{\nu' \mathbf{k}'}(\mathbf{r'}) \Psi_{\nu \mathbf{k}}(\mathbf{r'})    }{\omega - \epsilon_{\nu' \mathbf{k}'} + \epsilon_{\nu \mathbf{k}} + i \delta} \\
    &-  \frac{ \Psi_{\nu \mathbf{k}} (\mathbf{r}) \Psi^\dagger_{\nu' \mathbf{k}'}(\mathbf{r}) \Psi_{\nu' \mathbf{k}'}(\mathbf{r'}) \Psi^\dagger_{\nu \mathbf{k}}(\mathbf{r'})    }{\omega + \epsilon_{ \nu' \mathbf{k}'} - \epsilon_{\nu \mathbf{k}} - i \delta} \bigg] \ ,
\end{split}
\end{align}
where $\Psi_{\nu \mathbf{k} }$ and $\epsilon_{\nu \mathbf{k}}$ mark KS eigenstates and eigenvalues. 

\begin{figure}
\centering%
\includegraphics[width=0.5\linewidth]{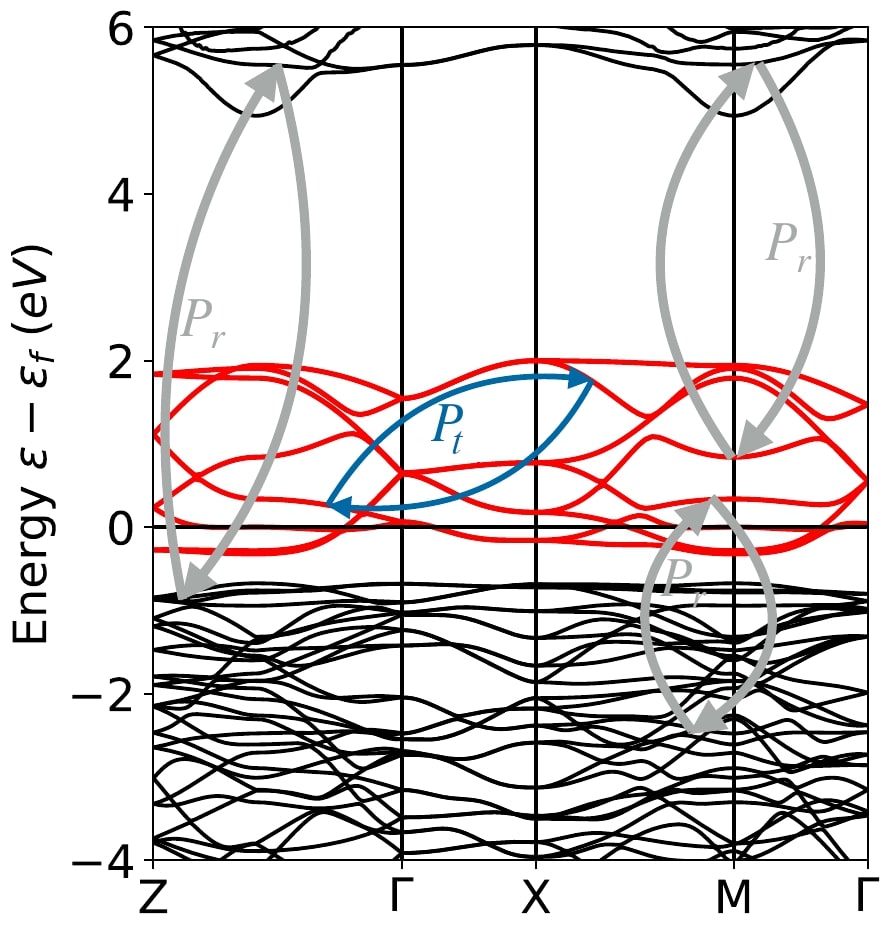}%
\caption{Band structure of LuNiO$_3$, with constructed Wannier functions for the Ni $e_g$ states as target states $t$. The decomposition in the polarization channels P$_{t}$ within the correlated subspace, and the polarization channels P$_r$ outside, from, and to the target subspace $t$ are schematically depicted as blue and grey arrows.}
\label{fig:crpa_example}
\end{figure}

The effective screened Coulomb interaction $U$ in a  target ``$(t)$'' space (typically the manifold of correlated states)can now be calculated as follows. Consider the example depicted in Fig.~\ref{fig:crpa_example}. The effective screened Coulomb interaction is calculated by first splitting the polarization of the system in two parts, $P_t$ pertaining only to transitions among the target orbitals and $P_r$ containing all other transitions (including from target to  non-target orbitals and the reverse):
\begin{align}
  P = P_{t} + P_r \quad .
  \label{eq:split_p}
\end{align}
Now one can deduce the partially screened interaction $W_r$ from $P_r$ as:
\begin{align}
  W_r = \epsilon^{-1}_r V = [ 1 - V \ P_r ]^{-1} V \ ,
\end{align}
Here $W_r$, which is implicitly restricted  to include only matrix elements among the target orbitals, is Coulomb interaction tensor $U_{m m' m'' m'''}$ for the target orbitals, screened by transitions involving other orbitals. Because the screening involves only the non-target orbitals it is referred to as ``constrained''. Adding the polarization $P_{t}$ to $P_r$ would recover the fully screened interaction
\begin{align}
  U = [ 1 - V \ P_r ]^{-1} V \quad .
\end{align}
Note, that due to the energy dependency of the polarization, $W_r$  is naturally frequency dependent. The frequency dependence  is often neglected and   $U(\omega=0)$ is used.

Now, one can analyze the symmetries of the calculated Coulomb tensor and fit to one of the two forms given above. Either by directly averaging the elements of the tensor, or using a minimization procedure to minimize differences between $U^{\text{cRPA}}$ and a constructed $U$. Importantly, $U$ should be evaluated in the same orbitals used for the embedding techniques. We advocate the use of a more advanced version of cRPA for disentangled bands implemented in VASP evaluating the polarization directly using the \acp{WF}~\cite{Merzuk2015}. 

It is evident that the screening depends crucially on the chosen subspace. For example, a larger energy window for the target space in the  downfolding  produces more atomic like orbitals but also has fewer screening channels, so leads to a  larger interaction, whereas smaller energy windows give more extended orbitals with smaller, heavily screened, interactions. As we will show later in Sec.~\ref{subsec:dmft_local_elec_structure} the screening in infinite layer nickelates is very sensitive as both, oxygen $p$ states below, and Nd $d$ states above, make large contributions to the screening. We discuss results from cRPA in Sec.~\ref{subsec:DMFT_low_energy}.

\subsection{Including local Coulomb interactions on oxygen: DFT+sicDMFT approach}
\label{sec:SIC_method}

In standard DFT+DMFT the many-body physics imposed by $U$ on Ni is treated within the DMFT correlated subspace, however the description of the quantum processes on O remains on the Kohn-Sham DFT level. Note that correlations on the O$(2p)$ orbitals are not necessarily weak because these orbitals, just as Ni$(3d)$ ones, carry the first new angular-momentum number with growing main quantum number (there is neither a $1p$ nor a $2d$ orbital), meaning the orbitals may sit close to the atomic nucleus and therefore are more compact and with a larger charging energy. This implies that also $2p$ frontier orbitals have a demanding pseudopotential that needs to moderate between localization and itinerancy (though not as severely as for $3d$ orbitals). 
And this nature becomes increasingly relevant when the connection to the environment via a $\Delta$ comparable to $U$ is substantial, a feature taking place for later TMOs. This suggests that a  DFT-level treatment may be insufficient to tackle the sophisticated Mott-Hubbard vs. charge-transfer balance. As a further aspect in infinite-layer nickelates, while Ni-$d_{x^2-y^2}$ is strongly hybridized with O$(2p)$, Ni-$d_{3z^2-r^2}$ is less so due to the missing apical oxygen. Hence the Ni-$e_g$ $\{3z^2-r^2,x^2-y^2\}$ orbitals of the DMFT-active Ni$(3d)$ shell are quite differently affected by O$(2p)$. This may also call for an improved description of correlation effects originating from oxygen. 

However, treating explicit Coulomb interaction within DMFT also for O$(2p)$ raises several methodological and numerical challenges, and the explicit quantum-fluctuating aspect in these orbitals should be still generally  weaker than in transition-metal $3d$ orbitals. In order to include localization tendencies on the oxygen sites beyond KS-DFT, one may therefore introduce the self-interaction correction (SIC)\cite{Perdew1981} as a simplified treatment compared to DMFT.  The SIC scheme can efficiently be applied already on the pseudopotential level~\cite{Vogel1996,Filippetti2003,Koerner2010}, leading to a refined O pseudopotential to be used in a standard CSC DFT+DMFT~\cite{Lechermann2019}. Use of this pseudopotential defines the DFT+sicDMFT method, which is thus capable of handling correlation physics on and between Ni and O closer to equal footing.

\begin{figure}
\centering
\includegraphics*[width=0.75\linewidth]{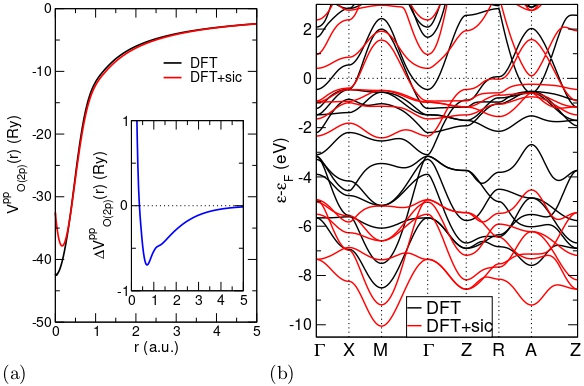}
\caption{Effects of SIC inclusion on the DFT level. (a) Radial part of oxygen $2p$ pseudopotential with (red) and without SIC (black). Inset: difference between DFT and DFT+sic pseudopotential.  (b) Band structure for NdNiO$_2$ within DFT (black) and DFT+sic (red). Note that in the latter calculation, SIC was only applied to O and the Ni orbitals are treated on the DFT level.}
\label{figsic0}
\end{figure}

Fig.~\ref{figsic0}a,b show the principal impact of the SIC inclusion on the DFT level. We see that the radial part of the O$(2p)$ pseudopotential is somewhat reduced within the lower-limit of the bonding region ($\sim$ 0.5-2 a.u.), resulting in a stronger localization of charge near the O site. For the NdNiO$_2$ crystal calculation within DFT+sic, i.e. employing the revised oxygen pseudopotential in the conventional KS cycle, two key effects may be observed. First, the O$(2p)$ block of six bands (originally between $\sim [-3.5,-8]$\,eV) get shifted to deeper energies, hence the $p-d$ splitting is increased. Second, especially around the Fermi level, some band-narrowing takes place due to the renormalized hoppings resulting from the increased charge localization. It is important to note however that here these bands are an intermediate step in the full DFT+sicDMFT scheme, not a final physical result and are shown to provide insight into the physics underlying the method. In particular, the upward shift of the self-doping band away from the Fermi level is an artifact of the neglect of the  local Coulomb interactions on Ni. The complete DFT+sicDMFT scheme yields the self-doping band again back at the Fermi level (cf. Fig.~\ref{figsic1}), which we will discuss in detail in Sec.~\ref{sec:lowdmft}.


\section{Results}
\label{sec:results}

In this section, we give an overview of results on infinite-layer nickelates in literature and a comparison to the better understood case of the layered copper oxides.  We mainly focus on theoretical results~\cite{Hu2019, Wang2020,Jiang2020,Si2020,Geisler2020,Sakakibara2020,He2020,Wu2020,Werner2020,Zhang2020,Wang2020b,Zhang2020c,Bernardini2020,Bernardini2020a,Liu2021,Wan2021,Plienbumrung2021,Malyi2021,Peng2021,Choubey2021,Kang2021,Sawatzky2019,Nomura2019, Nomura2020,Gu2020,Hirayama2020,Jin2020, Petocchi2020a, Higashi2021,Leonov2021,Lin2021a,Lee2004, Choi2020a, Choi2020, Karp:2020, Karp2020a, Karp:2021} but present some comparison to experiments~\cite{Li2019,Zeng2020,Li2020a,Hepting2020,Gu2020a,Goodge2021,Wang2021,Zhao2021,Lu2021,Osada2020,Osada2020a,Osada2021,Ren2021,Zeng2021,Puphal2021}, when the relevant experimental results are available. We also mention that while we are aware of the important works on the study of interface effects~\cite{Geisler2020, Bernardini2020a, He2020}, due to space limitation, we concentrate on the study of bulk nickelates here. 

We present four levels of results: for orientation we show the DFT-level electronic structure; then we describe the basic many-body electronic structure following from the different DFT+DMFT calculations and the approximate physical picture that results. Next we consider the predictions of these calculations for the Fermiology--which bands are present at the Fermi surface and what are the mass enhancements. Finally, inspired by recent experimental results, we make some brief remarks about magnetism in Sec.~\ref{subsec:magnetism_results}. 

\subsection{DFT Results}
\label{subsec:dft_results}

\begin{figure}
\centering
\includegraphics*[width=0.9\linewidth]{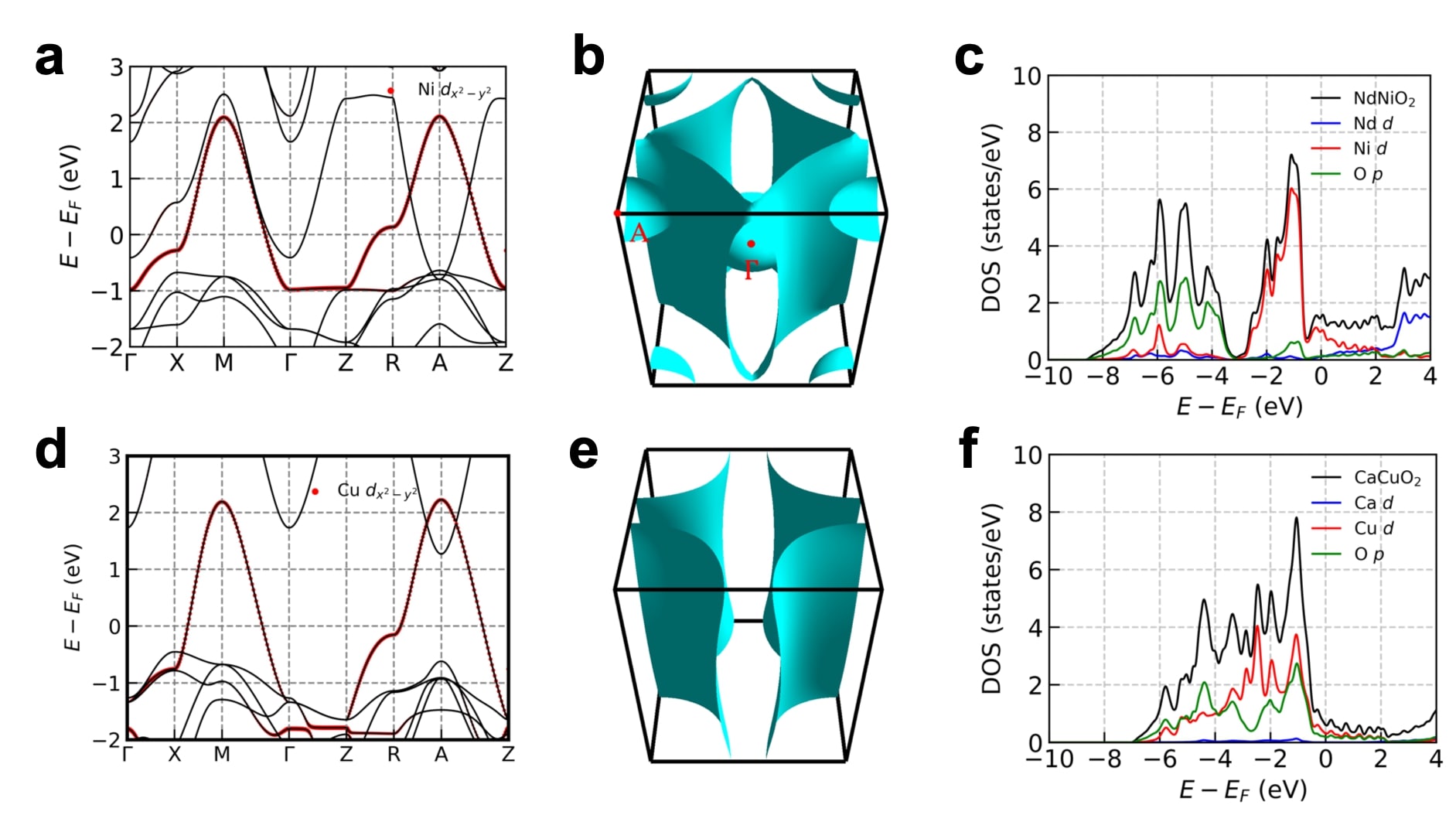}
\caption{\textbf{a}-\textbf{c}: Electronic properties of NdNiO$_2$. \textbf{a}: Electronic band structure of NdNiO$_2$ close to the Fermi level. The red dots highlight the Ni $d_{x^2-y^2}$ band. A second band also crosses the Fermi level. \textbf{b}: Fermi surface of NdNiO$_2$. In addition to the cylindrical Fermi sheet that is derived from Ni $d_{x^2-y^2}$ band, there are two additional electron pockets: one is at $\Gamma=(0,0,0)$ and the other is at $A=(\pi,\pi,\pi)$. \textbf{c}: Densities of states of NdNiO$_2$. The black, blue, red and green curves correspond to total, Nd-$d$ projected, Ni-$d$ projected and O-$p$ projected densities of states, respectively. The Fermi level is shifted to the zero point. \textbf{d}-\textbf{f}: Electronic properties of CaCuO$_2$. \textbf{d}: Electronic band structure of CaCuO$_2$ close to the Fermi level. The red dots highlight the Cu $d_{x^2-y^2}$ band. \textbf{e}: Fermi surface of CaCuO$_2$. \textbf{f}: Densities of states of CaCuO$_2$. The black, blue, red and green curves correspond to total, Ca-$d$ projected, Cu-$d$ projected and O-$p$ projected densities of states, respectively. The Fermi level is shifted to the zero point. Adapted from Ref.~{\cite{Gu2020}}.}
\label{figchen}
\end{figure}

Fig.~\ref{figchen} compares basic aspects of the \ac{DFT}-level electronic structure of the infinite layer nickelates NdNiO$_2$ and the analogous cuprate compound CaCuO$_2$. Panels \textbf{a} and \textbf{d} present the DFT bands of the two compounds. Panel \textbf{d} shows the familiar cuprate band structure, with one essentially two dimensional band of mixed Cu-$d_{x^2-y^2}$/O-$2p_\sigma$ character crossing the Fermi surface. Panel \textbf{a} shows that in NdNiO$_2$ the situation is richer, with other Fermi surface crossings in addition to  the  Ni-$d_{x^2-y^2}$-derived band (highlighted in red)~\cite{Lee2004,Gu2020,Botana2020,Liu2020,Kapeghian2020,Jiang2019}. The Fermi surfaces shown in panels \textbf{b} and \textbf{e} reveal that in addition to the $d_{x^2-y^2}$ bands which disperse very weakly in the $z$ direction  and give rise to a cylindrical Fermi surface sheet, there are two additional closed (three dimensional)  electron pockets centered at $\Gamma$ and $A$. The three dimensional sheets arise from  Nd-$d$ orbitals, in particular $d_{3z^2-r^2}$ and $d_{xy}$ orbitals, with an admixture of Ni-$d_{3z^2-r^2}$ and $d_{xz/yz}$ as well as interstitial states not directly attributed to any atomic orbital~\cite{Botana2020, Ryee2020, Jiang2019, Gu2020,Karp2020}.

Charge transfer from the $d_{x^2-y^2}$-derived band to the Nd-derived band leads to a ``self-doping" effect: in the stoichiometric infinite layer nickelate compound the $d_{x^2-y^2}$-derived band is not half-filled; rather its occupancy corresponds to about a 10-15$\%$ hole doping filling; thus stoichiometric nickelates should be compared to hole-doped cuprates. 

\begin{figure}
  \centering
  \includegraphics[width=0.9\linewidth]{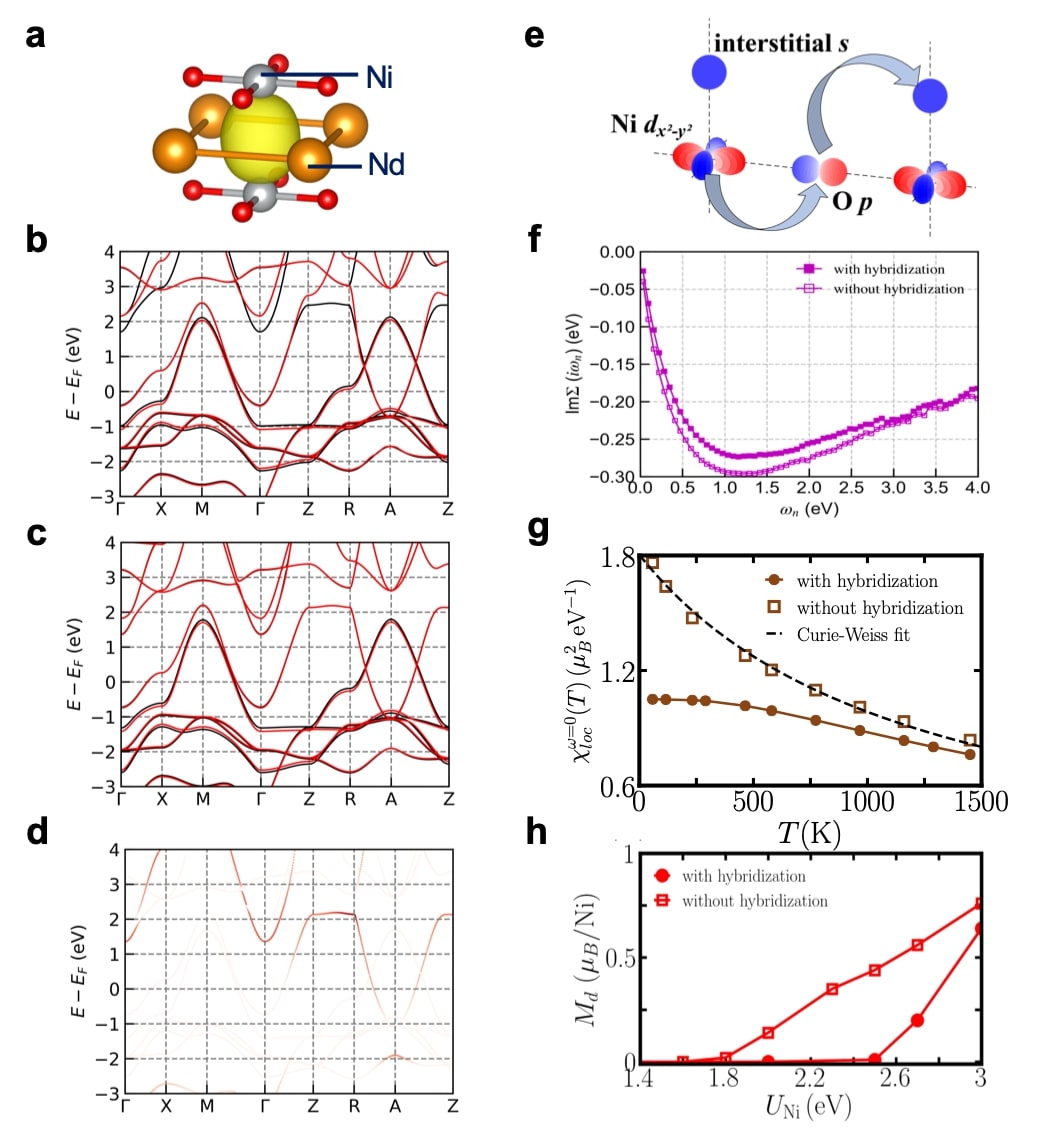}
  \caption{\textbf{a}: An iso-value surface of the interstitial $s$ orbital in NdNiO$_2$. \textbf{b}: Fitting of DFT band structure of NdNiO$_2$, using 16 Wannier functions (Nd-$d$, Ni-$d$ and O-$p$ orbitals). The black lines are DFT bands and the red lines are fitted bands from Wannier functions. \textbf{c}: Fitting of DFT band structure of NdNiO$_2$, using 17 Wannier functions (Nd-$d$, Ni-$d$, O-$p$ and the interstitial $s$ orbitals). The black and red lines have the same meaning as in panel \textbf{b}. \textbf{d}: The fatband plot of the interstitial $s$ orbital. \textbf{e}: An illustration of the hybridization between Ni $d_{x^2-y^2}$ orbital and the interstitital $s$ orbital via a second-nearest-neighbor hopping. \textbf{f}: Imaginary part of the self-energy of Ni $d_{x^2-y^2}$ orbital calculated using DFT+DMFT method with hybridization (solid symbols) and without hybridization (open symbols). \textbf{g}: Local susceptibility $\chi^{\omega=0}_{\textrm{loc}}(T)$ of NdNiO$_2$ as a function of temperature calculated with hybridization (solid symbols) and without hybridization (open symbols). The dashed line is a Curie-Weiss fitting. \textbf{h}: Magnetic moment on Ni $d$ orbitals as a function of interaction strength $\mathcal{U}_{\textrm{Ni}}$ calculated with hybridization (solid symbols) and without hybridization (open symbols). Adapted from Ref.~{\cite{Gu2020}}.}
\label{fig:interstitial}
\end{figure}

It is important to note that the published DFT analyses of orbital admixture are obtained by projecting the states onto atomic orbitals as described in the projector section above. Gu \textit{et al.}~\cite{Gu2020} find via a Wannier analysis that the additional band also has considerable contribution from charge density not centered on any atom. Because this component is not centered on an atom it is not easily revealed in the standard projector analysis. Panel \textbf{a} of Fig.~\ref{fig:interstitial} shows this component, known as an interstitial $s$ orbital, which is located at the mid-point between two neighboring Ni atoms along the $z$ axis. Panels \textbf{b} and \textbf{c} shows the fitting of DFT band structure using maximally localized Wannier functions (MLWF) as explained in the previous Methods section. Panel \textbf{b} shows a band fit based on 16 MLWFs, of which 5 are initialized as being centered on Ni-$d$ orbitals, 5 more centered on Nd-$d$ orbitals and 6 O-$p$ orbitals. The fitting is very good for the occupied bands but for one of the Nd-derived empty bands, the $Z \to R$ portion is not well reproduced. Panel \textbf{c} shows the result of adding one more MLWF that corresponds to the interstitial $s$ orbital. The fitting is improved, in particular in that Nd-derived bands are now exactly reproduced throughout the first Brillouin zone. Panel \textbf{d} shows the weight of the  interstitial $s$ orbital on the different bands. We can see that the $s$-orbital has a high weight on the extra band on the $Z \to R$ portion at energy $E\approx 2$ eV above the Fermi level.  

One of the key questions in the materials physics of the layered $d^9$ nickelates is whether the additional band is a ``spectator", acting simply as a reservoir enabling charge transfer from the NiO$_2$ plane to the Nd spacer layer, or whether the additional band also plays an essential role in the physics, either because it is strongly hybridized with or strongly interacting with the Ni degrees of freedom. The orbital composition of the additional band is relevant to this question: the ``spectator band" contains Nd $d_{3z^2-r^2}$ and $d_{xy}$ orbitals, as well as a small but non-zero admixture of Ni $d_{3z^2-r^2}$ and $d_{xz/yz}$ content; however the hybridization of these orbitals to the Ni $d_{x^2-y^2}$ band is very weak. However, in the Wannierization with the interstitial charge included, the hybridization between the interstitial $s$ orbital and Ni $d_{x^2-y^2}$ is one order of magnitude stronger ~\cite{Gu2020}. Panel \textbf{e} of Fig.~\ref{fig:interstitial} shows the hybridization between Ni $d_{x^2-y^2}$ and the interstitial $s$ orbital via the second-nearest-neighbor hopping. Because of this hopping, the itinerant electrons in the Nd spacer layer can effectively interact with the electrons in Ni $d_{x^2-y^2}$ orbital and therefore it is suggested that this coupling may lead to Kondo-type physics~\cite{Zhang2020a, Yang2021}. 

To complete the discussion of DFT-level theory we mention DFT+$U$ calculations on infinite-layer nickelates. In these calculations all the atomic orbitals that are in the pseudo-potentials are taken into account~\cite{Botana2020, Kapeghian2020, Been2021} and rotationally invariant Hubbard $U$ interactions are added on all the five Ni $d$ orbitals. Botana~\textit{et al.}~\cite{Botana2020} extracted various hopping matrices and energy splitting, which shows similarity between infinite-layer nickelates and cuprates. Kapeghian \textit{et al.}~\cite{Kapeghian2020}, Been~\textit{et al.}~\cite{Been2021} and Xia~\textit{et al.}~\cite{Xia2021} studied the electronic structure trends of the entire lanthanide series of infinite-layer nickelates.

\subsection{DFT+DMFT: local electronic structure}
\label{subsec:dmft_local_elec_structure}

In transition metal oxides, it is believed that the interesting correlation physics arises from a competition between local interactions within the transition metal $d$-shell, which control the relative energetics of different $d$-multiplets, and the hybridization with other orbitals, which acts to mix the $d$-multiplets. In assessing the relevance of different interactions, an analysis of the ground state wave function is of interest. As noted previously, a formal valence analysis places either the Ni or the Cu in a $d^9$ state, thus with one hole in the $d$-shell and full oxygen-$2p$ and empty Nd-$5d/6s$ shells. Deviations from this simple picture provide insight into the relevant interaction processes. One question is the admixture of ligand (O-${2p}$ holes in the cuprate and nickelate cases and also Nd-${5d/6s}$ electrons in the nickelate case) states. One distinguishes~\cite{Zaanen1985} ``charge transfer" materials where the energy difference between the ligand and transition metal $d$ states controls the physics from Mott Hubbard materials where the charging energy of the transition metal $d$-shells controls the physics. A second issue is the relative weight of different transition metal multiplets. In ``Hund's metals'', multiplet configurations involving high spin (spin $S\geq 1$) $d$-states are relevant; in Mott Hubbard materials only the $S=1/2$ and $S=0$ states are relevant.

DFT+DMFT calculations  provide theoretical estimates of orbital occupancies and of the local density matrices describing the multiplet probabilities of the correlated sites and the occupancies of the ligand sites.  In the cuprate case there is general agreement both within DFT and in DFT+DMFT that the only relevant states are $d^9$ (with the hole in the Cu-$d_{x^2-y^2}$ orbital) and $d^{10}\underbar{L}$. These two states appear with almost equal weight, while the Cu $d^8$ configuration plays a negligible role (for a recent calculation consistent with the substantial previous literature see \cite{Karp:2020}). This pattern of occupancies marks the cuprate material as a ``charge transfer" compound \cite{Zaanen1985} in which the major deviation from the atomic limit comes from moving the $d$-shell hole onto the oxygen network and back and the correlation physics should be thought of as arising from the oxygen-copper hybridization in the presence of strong local Cu correlations.  

In the nickelate compounds the theoretical situation is less clear. There is a general consensus that the Ni-$d_{x^2-y^2}$ orbital is occupied by approximately one electron, and that the oxygen states are farther removed in energy from the $d_{x^2-y^2}$ orbital than in the cuprates and also more weakly hybridized, implying the admixture of the oxygen states into the near Fermi level bands is smaller than in cuprates ~\cite{Sawatzky2019, Botana2020,Jiang2020,Karp2020a, Lechermann2020}. There is also general consensus that charge transfer onto the Nd states occurs. However, whether NdNiO$_2$ is in the Mott-Hubbard region or in a critical region with mixed charge-transfer/Mott-Hubbard character is still under debate~\cite{Sawatzky2019,Jiang2020,Shen2021}. Another complication arises from the interstitial $s$ orbital, which hybridizes with the Ni-$d_{x^2-y^2}$ orbital. Panels \textbf{f}, \textbf{g} and \textbf{h} of Fig.~\ref{fig:interstitial} compare the imaginary part of the self-energy of Ni-$d_{x^2-y^2}$ orbital, the local susceptibility and magnetic phase diagram of NdNiO$_2$ with the hybridization (solid symbols) and without the hybridization (open symbols)~\cite{Gu2020}, calculated using DFT+DMFT method ($U_{\textrm{Ni}} = 2$ eV) that is explained in the Methods section. Panel \textbf{f} shows $\textrm{Im}\Sigma(i\omega_n)$ of Ni-$d_{x^2-y^2}$ orbital. The effective mass $\frac{m^*}{m}\simeq 1-\frac{d \textrm{Im}(i\omega_n)}{d\omega_n}\rvert_{\omega_n\to0}$ is reduced from 2.0 without hybridization to 1.8 with hybridization. Panel \textbf{g} shows the local susceptibility $\chi^{\omega=0}_{\textrm{loc}}(T) = \int^{\beta}_0 \chi_{\textrm{loc}}(\tau)d\tau=\int^{\beta}_0g^2\langle S_z(\tau)S_z(0) \rangle  d\tau $. The hybridization reduces $\chi^{\omega=0}_{\textrm{loc}}(T)$ at low temperatures, indicating the screening of the Ni spin in $d_{x^2-y^2}$ orbital. Panel \textbf{h} shows the magnetic moment on Ni atom as a function of interaction strength $\mathcal{U}$ on Ni-$d_{x^2-y^2}$ orbital. The hybridization increases the critical $\mathcal{U}$ that is needed to stabilize long-range antiferromagnetic ordering. Overall, the presence of the hybridization makes Ni-$d_{x^2-y^2}$ orbital less correlated and less magnetic, which is consistent with the Kondo screening picture. 

Perhaps more importantly, unlike the cuprate case the transition metal Ni-$d_{3z^2-r^2}$ may also be relevant. The Ni-$d_{3z^2-r^2}$ has a non-negligible hybridization with the Nd-$d_{3z^2-r^2}$ band, so the three dimensional bands may be more than spectator bands, and instead participate to some degree in the correlation physics and Hund's physics may be relevant. Karp \textit{et al.} \cite{Karp2020,Karp2020a} find $\leq 15\%$ high spin $d^8$  and argue that only the Ni-$d_{x^2-y^2}$ orbital is important for the low-energy physics. Wang \textit{et al.} \cite{Wang2020b}  find $25.9\%$ high spin $d^8$ ($10.8\%$ low spin $d^8$) in the ground state configuration of LaNiO$_2$ and argue based on this that the material should be classified as a Hund's metal. A difference between the calculations is the number of correlated d orbitals retained.  The result of Wang \textit{et al.} \cite{Wang2020b} is in partial agreement with the GW+EDMFT study of \citeauthor{Petocchi2020a}~\cite{Petocchi2020a} which also finds $\sim 25\%$ high spin $d^8$ character ($\sim 25\%$ low spin $d^8$) at optimal doping level. 
However, \citeauthor{Petocchi2020a}~\cite{Petocchi2020a} report a nonmonotonic doping dependence of the  high-spin $d^8$ weight, whereas  Ref.~\cite{Wang2020b} reports a monotonic doping dependence. \citeauthor{Petocchi2020a}~\cite{Petocchi2020a} point out that the effect of this physics on the low energy properties is not clear.

These questions are not theoretically settled because, as shown by Karp \textit{et al.} \cite{Karp2021dependence} the choice of method used to construct the local orbitals of the downfolded model  affects the DMFT results for the nickelate.  Table \ref{tab:dmft_occ} compares the orbital occupancies and multiplet occurence probabilities (defined as weight of the different configurations in the many-body density matrix projected onto the Ni states) obtained with different methodologies.

\subsection{DMFT theory of the low energy physics\label{sec:lowdmft}}
\label{subsec:DMFT_low_energy}

\begin{figure}
\centering
\includegraphics*[width=0.95\linewidth]{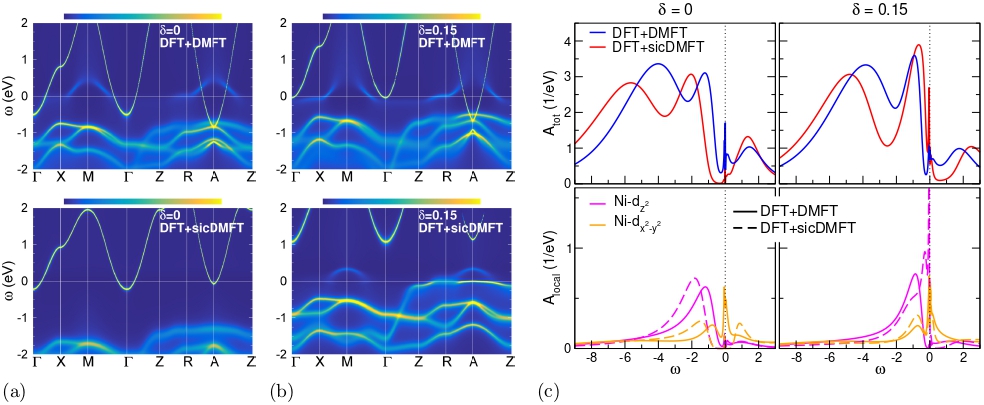}
\caption{Comparison between DFT+DMFT and DFT+sicDMFT at $T=30$\,K, using $U_{avg}=10$\,eV, $J_{avg}=1$\,eV and projected-local orbitals on the 12 KS states above the O$(2s)$ bands for both schemes, respectively.
(a,b) $k$-resolved spectral function $A({\bf k},\omega)$ for (a) pristine and and (b) $\delta=0.15$ hole-doped NdNiO$_2$. (c) Total and Ni-$e_g$ local spectral function for both doping cases. Adapted from~\cite{Lechermann2021}.}
\label{figsic1}
\end{figure}

We next consider the consequences of the wide window electronic structure for the low energy physics. Panels (a) and (b) of Fig.~\ref{figsic1} show the many body electronic structure (momentum and frequency dependent electron spectral function) computed for NdNiO$_2$ at two representative dopings using the basic DFT+DMFT and DFT+sicDMFT methods for a particular method and choice of parameters.
From a comparison with other nickelates~\cite{Lechermann2019} and in view of experimental constraints (see discussion in~\cite{Lechermann2020}) a value $U_{avg}=10$\,eV and $J_{avg}=1$\,eV is here used to parametrize the local Coulomb interaction for the Ni$(3d)$ orbitals in charge self-consistent DFT+(sic)DMFT calculations. The projector method for the five Ni$(3d)$ orbitals, building on the 12 KS states states above the O$(2s)$ bands (i.e. an energy window ~[-10,3]\,eV), is employed and a rotational-invariant Slater Hamiltonian is active in the resulting correlated subspace. The hole doping $\delta$ is achieved by the virtual-crystal approximation using an
effective Nd atom~\cite{Lechermann2021}, where the Nd$(4f)$ states are frozen
in the pseudopotential core. Note again that calculational settings resulting in the data shown in Fig~\ref{figsic1} differ only in using the LDA(SIC) oxygen
pseudpotential in DFT+(sic)DMFT.

The upper panels of Figs~\ref{figsic1}(a,b) show the DFT+DMFT spectral function. In comparison to the DFT bands shown in Fig. ~\ref{figchen}, we see that the Ni-$d_{x^2-y^2}$-derived band is narrowed and broadened, and its separation in energy from the lower lying Ni-$t_{2g}$ and O$(2p)$ bands is increased. The lower panels show the DFT+sicDMFT results, which are markedly different. We see that in the stoichiometric compound the $d_{x^2-y^2}$ bands are completely absent: within this scheme these orbitals are completely localized and incoherent, hence not visible in the spectral function. The position of the `spectator' bands relative to the Fermi surface is also changed, with some electrons transferred to these orbitals~\cite{Lechermann2020a}. This may calls for an
alternative possible Kondo scenario at low $T$, including also a substantial role of the Ni-$d_{3z^2-r^2}$
orbital~\cite{Lechermann2020a}.
At doping $\delta=0.15$ we see that in the DFT+sicDMFT calculation some aspects of the $d_{x^2-y^2}$ bands are restored, but again the relative positions of the spectator and $d_{x^2-y^2}$ bands are very different in the two methods. Panel (c) of  Fig.~\ref{figsic1} shows the momentum integrated total and Ni-projected spectral functions. Important differences between the results of the two methods include the $p-d$ splitting (visible as the shift in higher binding energy peak from $\sim -4$\,eV to $\sim -6$\,eV (cf upper left panel) and the proximity of the $d_{3z^2-r^2}$ states to the Fermi surface. This characterizes the material as an effective orbital-selective Mott-insulator. In this theory, hole doping leaves the  Ni-$d_{x^2-y^2}$ occupancy almost unchanged at half filling. Instead, the Ni-$d_{3z^2-r^2}$ orbital takes care of most of the charge doping and becomes significantly further depleted. Therefore within DFT+sicDMFT, the superconducting region is designated by the coexistence of nearly half-filled Ni-$d_{x^2-y^2}$ and a Ni-$d_{3z^2-r^2}$-based flat band crossing the Fermi level. As discussed in Ref.~\cite{Lechermann2021}, this flat band interacts with the Mott-like state such as to increase coherency within the Ni-$d_{x^2-y^2}$ sector. For even larger hole doping, the system evolves into a bad Hund metal, where coherence is lost again~\cite{Lechermann2021}. Let us note that the shift of the Ni-$d_{3z^2-r^2}$-based flat band towards the Fermi level is supported by a full GW+EDMFT investigation~\cite{Petocchi2020a}.

\begin{figure}
\centering
\includegraphics*[width=0.75\linewidth]{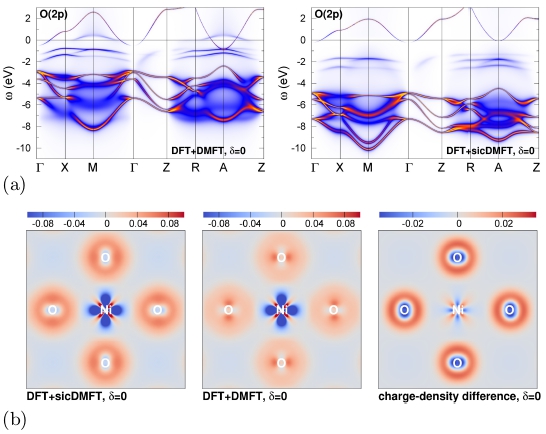}
\caption{Oxygen $2p$ states in DFT+DMFT and DFT+sicDMFT. (a) Orbital weight (i.e. fatbands) in the interacting regime along high symmetry lines. (b) Interacting bond charge density $\rho-\rho_{\rm atom}^{\rm LDA}$, with right panel displaying the difference between the densities shown in the left and middle panel.}
\label{figsic2}
\end{figure}

As depicted in Fig.~\ref{figsic2}, the increase of correlation strength with SIC inclusion originates from the stronger localization of O($2p$) electrons. The oxygen states are shifted down in energy, leading to an increase of the $p$-$d$ splitting and thus to a value $\Delta=5$\,eV for the charge-transfer energy~\cite{Lechermann2020}. Fig.~\ref{figsic2}b shows directly in real space, that the SIC-modified pseudopotential of oxygen enhances the $2p$ charge density around the oxygen sites, with additionally further depleting Ni-$d_{x^2-y^2}$. Hence in DFT+sicDMFT an adjustment of the $U$ vs. $\Delta$ competition takes place, which refines furthermore the various hopping integrals of the system.

These results highlight the basic electronic structure questions:  (i) what is the coherence (scattering rate and mass enhancement) of the Ni-$d_{x^2-y^2}$-derived band and (ii) how strong are the interactions on the `spectator' bands? (iii) what is the energy difference between the O-${2p}$ and $d_{x^2-y^2}$ orbitals. The different implementations of the DFT+DMFT methodology give different answers to these questions. While most methods, with the exception of the sic method, give a somewhat coherent $d_{x^2-y^2}$ band the estimates of the orbital mass enhancement vary substantially, as shown in Table~\ref{tab:mass_enhancements}. These quantities are experimentally accessible via photoemission experiments, and future experiment/theory comparisons will provide valuable methodological guidance. It should be noted that most papers only discuss orbital mass enhancements, which are generally larger than band enhancements, especially for large energy window calculations.

The question of band renormalization in different downfolding choices is tightly coupled with the question of screening (see Sec.~\ref{sec:crpa}). cRPA studies showed that if a 7 orbital downfolding model is constructed from five Ni$(3d)$ orbitals and two Nd$(5d)$ orbitals in a small energy window (not containing any oxygen states), the static part of the onsite Coulomb interaction is $U_{d_{x^2-y^2}}\approx 5$~eV~\cite{Nomura2019, Petocchi2020a}. For a minimal model containing only the Ni $d_{x^2-y^2}$, the Nd $d_{3z^2-r^2}$, and the interstitial s orbital, $U_{d_{x^2-y^2}}$ is even further reduced to $\approx 3.1$~eV~\cite{Nomura2019}. The bare Coulomb interaction is of the order of $V_{d_{x^2-y^2}}\approx 25$~eV, highlighting the strong screening of the low-energy states. The resulting strong frequency dependency, which is usually neglected in DFT+DMFT calculations, plays a crucial role leading to a mass enhancement in GW+EDMFT calculations comparable to that of DFT+DMFT calculations with much larger static $\mathcal{U}$ values (see Table~\ref{tab:mass_enhancements})~\cite{Petocchi2020a}. This shows, that the other orbitals close to the Fermi level play an important role in screening processes for the Ni $d_{x ^2-y^2}$ correlations, and using static Coulomb interaction parameters from cRPA will lead to a underestimation of correlation effects in a small energy window downfolding scheme. This also raises the importance of inter-site interaction Coulomb matrix elements in a large energy window calculation between the Ni$(3d)$ orbitals and O$(2p)$ and Nd$(5d)$ states. 

\begin{figure}
  \centering
  \includegraphics[width=\linewidth]{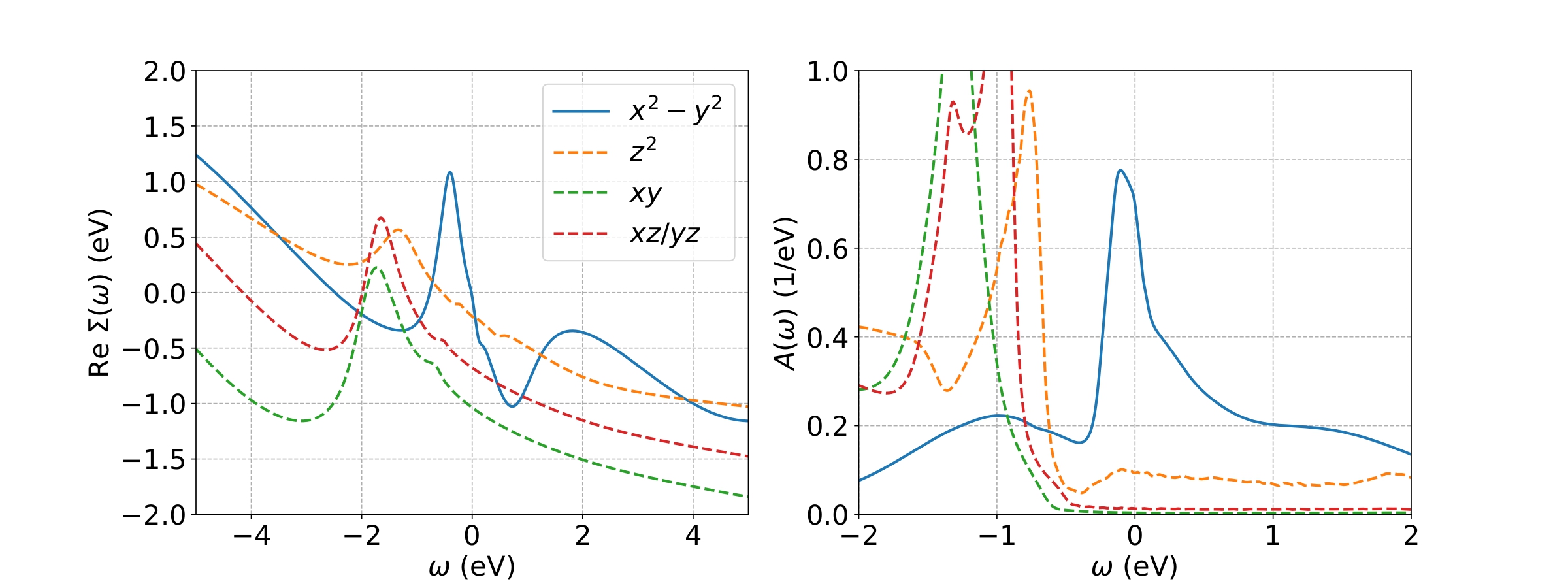}
  \caption{Left: Self energy of the different orbitals in a five orbital fully charge self consistent DFT+DMFT calculation at stoichiometry with projectors in an energy range from -10 eV to 10 eV around the Fermi level, using a rotationally invariant slater Hamiltonian with $U_{avg} = 7 eV$,  $J_{avg} = 0.7 eV$, and $T = 290K$. Right: corresponding momentum integrated spectral function. Adapted from Ref.~\cite{Karp2020a}.}
\label{fig:five_Sigma_A_w}
\end{figure}

Finally, we may consider the functional form of the self energy, which is relevant to the issue of Hund's physics. Hund's metal physics is believed to imply a  particle-hole asymmetric structure in the self energy leading to  an extra peak in the electron spectral function~\cite{Stadler19,Karp2020b}; for materials such as NdNiO$_2$ where the $d$-shell is more than half filled the peak is on the unoccupied part of the spectrum \cite{Karp2020b}. Conversely, Mott physics would result in a two peak spectral function and a more symmetric self energy with peaks on both sides.  
As shown in Fig.~\ref{fig:five_Sigma_A_w}, the $d_{x^2-y^2}$ self energy computed in ~\cite{Karp2020a} has strong (but broadened, especially on the positive frequency side)  peaks at $\omega\approx -0.7eV$ and $\omega\approx 2eV$ and the corresponding  spectral function shows two Hubbard peaks and a central quasiparticle feature. These are consistent with expectations of a Mott-Hubbard material. In addition, a weak feature is visible as a change of concavity around $\omega=0.2eV$; comparison to spectra presented for the Hund's metals Sr$_2$RuO$_4$ and Sr$_2$MoO$_4$ suggests that this may be a signature of weak Hund's physics~\cite{Karp:2020}. Wang \textit{et al.} report a much larger $d^8$ weight in the ground state but exhibit  a similar $d_{x ^2-y^2}$ spectral function, suggesting that the presence of some admixture of high-spin $d^8$ does not strongly affect the low energy physics. \citeauthor{kang2020infinitelayer}~\cite{kang2020infinitelayer} present a spectral function that shows a stronger feature (a  peak) at about $\omega=0.2eV$ on the unoccupied side of the spectral function at low temperatures, which may be a sign of Hund's correlations. They also report a slope inversion on the occupied side which is not expected in the theory of Hund's metals with more than half filled d-shells \cite{Karp:2020}. 
 


\subsection{Magnetism}
\label{subsec:magnetism_results}

We now discuss briefly the magnetic properties. The stoichiometric cuprates are antiferromagnetic insulators, with a charge gap of approximately $1.5$\,eV, a Neel temperature $\approx 300$\,K set by weak interlayer and spin orbit effects, and an intrinsically large magnetic scale evident for example as a zone boundary magnon energy $\approx 0.3$\,eV. Fitting to a Heisenberg model implies a nearest neighbor exchange coupling $J_{\text{NN}} \approx 120$ meV although it must be born in mind that the materials are intermediate coupling, so the magnetism is in a intermediate regime between itinerant and localized. Upon doping the commensurate magnetism vanishes rapidly but evidence of strong magnetic correlations (and in some materials tendency to incommensurate magnetic order) persists to a doping of about $0.15$.

A discussion of magnetism in the nickelates is complicated by the self-doping effect. If the physics of the nickelates is directly comparable to that of the cuprates then one would expect the stoichiometric nickelates to be on the boundary of magnetism. Recent resonant inelastic X-ray (RIXS) \cite{Lu2021} and nuclear magnetic resonance experiments ~\cite{Zhao2021} are consistent with this picture.  In particular, Lu \textit{et al.} report no long ranged order but observe a strong zone boundary paramagnon-like excitation implying a $J_{\text{NN}} \approx 60$ meV, about half of the cuprate value.

Theoretically a number of DFT+$U$, hybrid DFT and DFT+DMFT studies have studied magnetism of infinite layer nickelates~\cite{Botana2020, Wan2021, Zhang2021, Liu2020, Gu2020, Karp2020, Ryee2020, Choi2020a, Lechermann2021} and reported a wide range of magnetic superexchange~\cite{Jiang2020, Liu2020, Zhang2020, Katukuri2020, Wan2021, Wu2020, Nomura2020,Hirayama2020} from a small value of about 10 meV~\cite{Jiang2020,Liu2020} to an intermediate value of about 30 meV~\cite{Zhang2020} to a large value of about 80-100 meV~\cite{Katukuri2020,Wan2021,Wu2020,Nomura2020,Hirayama2020}. DFT+DMFT studies have not examined the Ni exchange coupling systematically. Further investigation of the magnetic properties within this method is likely to provide valuable insights.

\section{Summary}
DFT and beyond DFT analyses have produced a broadly coherent picture of the electronic structure of the infinite layer $d^9$ nickelates. Similar to the cuprates there is a rather two dimensional Ni-$d_{x^2-y^2}$-derived band that is moderately to strongly correlated. Differently from the cuprates, at the Fermi surface there is an additional, much more three dimensional, band derived from the rare earth $d$-orbitals (with some admixture of interstitial charge and of Ni$(3d)$ states). The energy difference between Ni$(3d)$ and O$(2p)$ orbitals is larger in the nickelates than the cuprates, putting the nickelates farther from the charge transfer regime than are the cuprates. Finally, in contrast to the cuprate materials where the only relevant configurations of the transition metal ions are the $d^9$ and $d^{10}$ states, in the nickelate materials some admixture of the high-spin $d^8$ configuration occurs, raising the possibility of Hund's metal physics. 

Given this broad consensus, the question becomes which of the differences and similarities to the cuprates are important for the low energy physics. It is clear that the additional band ``self-dopes'' the Ni $d_{x^2-y^2}$ bands, so that the chemistry-doping phase diagram of the  infinite layer nickelates is shifted from that of the cuprates by about $0.1$ hole/Ni, and that charge transfer to the oxygen orbitals  is less relevant in the nickelates than in the cuprates.  The important open question is whether  the other differences are important for the low energy correlation physics, in other words, whether the low energy physics of the infinite layer nickelates may be understood in terms of a one band Hubbard model or whether richer physics is needed. Answering this question bears directly on the issue of the mechanism for the observed superconductivity.  
This question is not yet settled, in part because different flavors of the \ac{DFT}+\ac{DMFT} have provided different quantitative answers to questions including the fractional weight of high spin $d^8$ configurations in the ground state, the relative energy positions of the $p$ and $d$ band manifolds and what is the mass enhancement of the different bands near Fermi surface. Some of these differences may be traced to different choices required in the DFT+DMFT approach to correlated materials. 

In this article we have explained the different choices and the different results that emerge. Some of the differences in results are experimentally testable. For example, the DFT+(sic)DMFT approach yield significantly more strongly correlated/less coherent $d_{x^2-y^2}$ derived bands and indeed different DFT+DMFT methods produce different mass enhancements. Thus angle-resolved photoemission measurements of the quasiparticle dispersion and linewidth, in combination with higher energy measurements of the $p$-$d$ energy splitting, can experimentally test the different predictions. Hund's metal physics is an intrinsically multiband effect, which involves characteristic asymmetries in the electron self energy. Even more importantly the known examples of Hund's metals involve multiple strongly correlated bands crossing the Fermi surface. Detailed analyses of the structure of the electron dispersion and the strength of the correlations on the ``spectator" bands will provide insight. 
There are also interesting differences in the theoretical predictions for magnetic properties and superconducting pairing~\cite{Wu2020,Kitatani2020,Wang2020a,Adhikary2020,Sakakibara2020,Zhang2020b,Gu2020a}, which future experiments may test. Finally, we note that our discussion is based on the single-site version of dynamical mean field theory. While model-system studies have shown that this approximation captures the main features of the wide energy range many-body electronic structure, the single-site approximation does not capture important aspects of the low energy physics, such as  superconductivity or pseudogap physics. ``Cluster" and related extensions (e.g. D$\Gamma$A) of the theory are an important directions for future research~\cite{Kitatani2020}. 

We hope that our work  will motivate comparisons to experiment and to more fundamental theoretical approaches that will help resolve some of the methodological questions relating to the DFT+DMFT approach to correlated materials.


\section*{Conflict of Interest Statement}

The authors declare that the research was conducted in the absence of any commercial or financial relationships that could be construed as a potential conflict of interest.

\section*{Author Contributions}

All authors contributed to the planning and writing of this paper, which is a review and does not report original research. 

\section*{Funding}
H.C. is supported by the National Natural Science Foundation of China under project number 11774236, the Ministry of Science and Technology of China under project number SQ2020YFE010418 and NYU University Research Challenge Fund. F. L. is supported by the Flatiron Institute. The Flatiron Institute is a division of the Simons Foundation. 

\section*{Acknowledgments}
 H.C. acknowledges useful discussion with Yuhao Gu, Mi Jiang, Danfeng Li, Jiawei Mei, Xiangang Wan, Tao Wu, Xianxin Wu, Yifeng Yang, Guangming Zhang, Zhicheng Zhong, Wei Zhu and in particular, Chengliang Xia for his help in making part of the figures. NYU high performance computing at Shanghai, New York and Abu Dhabi campuses provides the computational resources. Computations were partly performed at the University of Hamburg and the JUWELS Cluster of the J\"ulich Supercomputing Centre (JSC) under project number hhh08.



\bibliographystyle{frontiersinHLTH&FPHY} 

\begin{thebibliography}{119}
\expandafter\ifx\csname natexlab\endcsname\relax\def\natexlab#1{#1}\fi
\expandafter\ifx\csname urlstyle\endcsname\relax
  \expandafter\ifx\csname doi\endcsname\relax
  \def\doi#1{doi:\discretionary{}{}{}#1}\fi \else
  \expandafter\ifx\csname doi\endcsname\relax
  \def\doi{doi:\discretionary{}{}{}\begingroup \urlstyle{rm}\Url}\fi \fi
\expandafter\ifx\csname selectlanguage\endcsname\relax
  \def\selectlanguage#1{}\fi

\bibitem[{Bednorz and M{\"u}ller(1986)}]{Bednorz86}
Bednorz JG, M{\"u}ller KA.
\newblock {Possible high T$_c$ superconductivity in the Ba-La-Cu-O system}.
\newblock {\em Z. Phys. B Condensed Matter\/} {\bf 64} (1986) 189--193.
\newblock \doi{https://doi.org/10.1007/BF01303701}.

\bibitem[{Wu et~al.(1987)Wu, Ashburn, Torng, Hor, Meng, Gao et~al.}]{Chu87}
Wu MK, Ashburn JR, Torng CJ, Hor PH, Meng RL, Gao L, et~al.
\newblock Superconductivity at 93 k in a new mixed-phase y-ba-cu-o compound
  system at ambient pressure.
\newblock {\em Phys. Rev. Lett.\/} {\bf 58} (1987) 908--910.
\newblock \doi{10.1103/PhysRevLett.58.908}.

\bibitem[{Orenstein and Millis(2000)}]{Millis00}
Orenstein J, Millis AJ.
\newblock Advances in the physics of high-temperature superconductivity.
\newblock {\em Science\/} {\bf 288} (2000) 468--474.
\newblock \doi{10.1126/science.288.5465.468}.

\bibitem[{Li et~al.(2019)Li, Lee, Wang, Osada, Crossley, Lee et~al.}]{Li2019}
Li D, Lee K, Wang BY, Osada M, Crossley S, Lee HR, et~al.
\newblock {Superconductivity in an infinite-layer nickelate}.
\newblock {\em Nature\/} {\bf 572} (2019) 624--627.
\newblock \doi{10.1038/s41586-019-1496-5}.

\bibitem[{Anisimov et~al.(1999)Anisimov, Bukhvalov, and Rice}]{Anisimov99}
Anisimov VI, Bukhvalov D, Rice TM.
\newblock Electronic structure of possible nickelate analogs to the cuprates.
\newblock {\em Phys. Rev. B\/} {\bf 59} (1999) 7901--7906.
\newblock \doi{10.1103/PhysRevB.59.7901}.

\bibitem[{LeBlanc et~al.(2015)LeBlanc, Antipov, Becca, Bulik, Chan, Chung
  et~al.}]{simonscollab:2015}
LeBlanc JPF, Antipov AE, Becca F, Bulik IW, Chan GKL, Chung CM, et~al.
\newblock Solutions of the two-dimensional hubbard model: Benchmarks and
  results from a wide range of numerical algorithms.
\newblock {\em Phys. Rev. X\/} {\bf 5} (2015) 041041.
\newblock \doi{10.1103/PhysRevX.5.041041}.

\bibitem[{Zheng et~al.(2017)Zheng, Chung, Corboz, Ehlers, Qin, Noack
  et~al.}]{Zheng17}
Zheng BX, Chung CM, Corboz P, Ehlers G, Qin MP, Noack RM, et~al.
\newblock Stripe order in the underdoped region of the two-dimensional hubbard
  model.
\newblock {\em Science\/} {\bf 358} (2017) 1155--1160.
\newblock \doi{10.1126/science.aam7127}.

\bibitem[{Jiang and Devereaux(2019)}]{Jiang19}
Jiang HC, Devereaux TP.
\newblock Superconductivity in the doped hubbard model and its interplay with
  next-nearest hopping $t'$.
\newblock {\em Science\/} {\bf 365} (2019) 1424--1428.
\newblock \doi{10.1126/science.aal5304}.

\bibitem[{Kitatani et~al.(2020)Kitatani, Si, Janson, Arita, Zhong, and
  Held}]{Kitatani2020}
Kitatani M, Si L, Janson O, Arita R, Zhong Z, Held K.
\newblock {Nickelate superconductors—a renaissance of the one-band Hubbard
  model}.
\newblock {\em npj Quantum Materials\/} {\bf 5} (2020) 59.
\newblock \doi{10.1038/s41535-020-00260-y}.

\bibitem[{Hepting et~al.(2020)Hepting, Li, Jia, Lu, Paris, Tseng
  et~al.}]{Hepting2020}
Hepting M, Li D, Jia CJ, Lu H, Paris E, Tseng Y, et~al.
\newblock {Electronic structure of the parent compound of superconducting
  infinite-layer nickelates}.
\newblock {\em Nature Materials\/} {\bf 19} (2020) 381--385.
\newblock \doi{10.1038/s41563-019-0585-z}.

\bibitem[{Been et~al.(2021)Been, Lee, Hwang, Cui, Zaanen, Devereaux
  et~al.}]{Been2021}
Been E, Lee WS, Hwang HY, Cui Y, Zaanen J, Devereaux T, et~al.
\newblock {Electronic Structure Trends Across the Rare-Earth Series in
  Superconducting Infinite-Layer Nickelates}.
\newblock {\em Physical Review X\/} {\bf 11} (2021) 011050.
\newblock \doi{10.1103/PhysRevX.11.011050}.

\bibitem[{Hu and Wu(2019)}]{Hu2019}
Hu LH, Wu C.
\newblock {Two-band model for magnetism and superconductivity in nickelates}.
\newblock {\em Physical Review Research\/} {\bf 1} (2019) 032046.
\newblock \doi{10.1103/PhysRevResearch.1.032046}.

\bibitem[{Werner and Hoshino(2020)}]{Werner2020}
Werner P, Hoshino S.
\newblock {Nickelate superconductors: Multiorbital nature and spin freezing}.
\newblock {\em Physical Review B\/} {\bf 101} (2020) 041104.
\newblock \doi{10.1103/PhysRevB.101.041104}.

\bibitem[{Zhang and Vishwanath(2020)}]{Zhang2020b}
Zhang YH, Vishwanath A.
\newblock {Type-II $t-J$ model in superconducting nickelate
  Nd$_{1-x}$Sr$_x$NiO$_2$}.
\newblock {\em Physical Review Research\/} {\bf 2} (2020) 023112.
\newblock \doi{10.1103/PhysRevResearch.2.023112}.

\bibitem[{Kang and Kotliar(2021)}]{Kang2021}
Kang CJ, Kotliar G.
\newblock {Optical Properties of the Infinite-Layer La$_{1-x}$Sr$_x$NiO$_2$ and
  Hidden Hund's Physics}.
\newblock {\em Physical Review Letters\/} {\bf 126} (2021) 127401.
\newblock \doi{10.1103/PhysRevLett.126.127401}.

\bibitem[{Wang et~al.(2020{\natexlab{a}})Wang, Kang, Miao, and
  Kotliar}]{Wang2020b}
Wang Y, Kang CJ, Miao H, Kotliar G.
\newblock {Hund's metal physics: From SrNiO$_2$ to LaNiO$_2$} {\bf 102}
  (2020{\natexlab{a}}) 161118.
\newblock \doi{10.1103/PhysRevB.102.161118}.

\bibitem[{Wan et~al.(2021)Wan, Ivanov, Resta, Leonov, and Savrasov}]{Wan2021}
Wan X, Ivanov V, Resta G, Leonov I, Savrasov SY.
\newblock {Exchange interactions and sensitivity of the Ni two-hole spin state
  to Hund's coupling in doped NdNiO$_2$}.
\newblock {\em Physical Review B\/} {\bf 103} (2021) 075123.
\newblock \doi{10.1103/PhysRevB.103.075123}.

\bibitem[{Wu et~al.(2020)Wu, {Di Sante}, Schwemmer, Hanke, Hwang, Raghu
  et~al.}]{Wu2020}
Wu X, {Di Sante} D, Schwemmer T, Hanke W, Hwang HY, Raghu S, et~al.
\newblock {Robust $d_{x^2-y^2}$-wave superconductivity of infinite-layer
  nickelates}.
\newblock {\em Physical Review B\/} {\bf 101} (2020) 060504.
\newblock \doi{10.1103/PhysRevB.101.060504}.

\bibitem[{Nomura et~al.(2019)Nomura, Hirayama, Tadano, Yoshimoto, Nakamura, and
  Arita}]{Nomura2019}
Nomura Y, Hirayama M, Tadano T, Yoshimoto Y, Nakamura K, Arita R.
\newblock {Formation of a two-dimensional single-component correlated electron
  system and band engineering in the nickelate superconductor NdNiO$_2$}.
\newblock {\em Physical Review B\/} {\bf 100} (2019) 205138.
\newblock \doi{10.1103/PhysRevB.100.205138}.

\bibitem[{Lechermann(2020{\natexlab{a}})}]{Lechermann2020a}
Lechermann F.
\newblock {Multiorbital Processes Rule the Nd$_{1-x}$Sr$_x$NiO$_2$ Normal
  State}.
\newblock {\em Physical Review X\/} {\bf 10} (2020{\natexlab{a}}) 041002.
\newblock \doi{10.1103/PhysRevX.10.041002}.

\bibitem[{Gu et~al.(2020{\natexlab{a}})Gu, Zhu, Wang, Hu, and Chen}]{Gu2020}
Gu Y, Zhu S, Wang X, Hu J, Chen H.
\newblock {A substantial hybridization between correlated Ni-$d$ orbital and
  itinerant electrons in infinite-layer nickelates}.
\newblock {\em Communications Physics\/} {\bf 3} (2020{\natexlab{a}}) 84.
\newblock \doi{10.1038/s42005-020-0347-x}.

\bibitem[{Gao et~al.(2021)Gao, Peng, Wang, Fang, and Weng}]{Gao2021}
Gao J, Peng S, Wang Z, Fang C, Weng H.
\newblock {Electronic structures and topological properties in nickelates
  $Ln_{n +1}$Ni$_n$O$_{2n +2}$}.
\newblock {\em National Science Review\/} {\bf 8} (2021).
\newblock \doi{10.1093/nsr/nwaa218}.

\bibitem[{Jiang et~al.(2020)Jiang, Berciu, and Sawatzky}]{Jiang2020}
Jiang M, Berciu M, Sawatzky GA.
\newblock {Critical Nature of the Ni Spin State in Doped NdNiO$_2$}.
\newblock {\em Physical Review Letters\/} {\bf 124} (2020) 207004.
\newblock \doi{10.1103/PhysRevLett.124.207004}.

\bibitem[{Botana and Norman(2020)}]{Botana2020}
Botana AS, Norman MR.
\newblock {Similarities and Differences between LaNiO$_2$ and CaCuO$_2$ and
  Implications for Superconductivity}.
\newblock {\em Physical Review X\/} {\bf 10} (2020) 011024.
\newblock \doi{10.1103/PhysRevX.10.011024}.

\bibitem[{Karp et~al.(2020{\natexlab{a}})Karp, Botana, Norman, Park, Zingl, and
  Millis}]{Karp2020}
Karp J, Botana AS, Norman MR, Park H, Zingl M, Millis A.
\newblock {Many-Body Electronic Structure of NdNiO $_2$ and CaCuO$_2$}.
\newblock {\em Physical Review X\/} {\bf 10} (2020{\natexlab{a}}) 021061.
\newblock \doi{10.1103/PhysRevX.10.021061}.

\bibitem[{Kapeghian and Botana(2020)}]{Kapeghian2020}
Kapeghian J, Botana AS.
\newblock {Electronic structure and magnetism in infinite-layer nickelates
  $R$NiO$_2$ ($R$ = La-Lu)}.
\newblock {\em Physical Review B\/} {\bf 102} (2020) 205130.
\newblock \doi{10.1103/PhysRevB.102.205130}.

\bibitem[{Karp et~al.(2020{\natexlab{b}})Karp, Hampel, Zingl, Botana, Park,
  Norman et~al.}]{Karp2020a}
Karp J, Hampel A, Zingl M, Botana AS, Park H, Norman MR, et~al.
\newblock {Comparative many-body study of Pr 4 Ni 3 O 8 and NdNiO 2}.
\newblock {\em Physical Review B\/} {\bf 102} (2020{\natexlab{b}}) 245130.
\newblock \doi{10.1103/PhysRevB.102.245130}.

\bibitem[{Park et~al.(2012)Park, Millis, and Marianetti}]{Park12}
Park H, Millis AJ, Marianetti CA.
\newblock Site-selective mott transition in rare-earth-element nickelates.
\newblock {\em Phys. Rev. Lett.\/} {\bf 109} (2012) 156402.
\newblock \doi{10.1103/PhysRevLett.109.156402}.

\bibitem[{Mizokawa et~al.(2000)Mizokawa, Khomskii, and Sawatzky}]{Mizokawa00}
Mizokawa T, Khomskii DI, Sawatzky GA.
\newblock Spin and charge ordering in self-doped mott insulators.
\newblock {\em Phys. Rev. B\/} {\bf 61} (2000) 11263--11266.
\newblock \doi{10.1103/PhysRevB.61.11263}.

\bibitem[{Georges(2004)}]{Georges04}
Georges A.
\newblock Strongly correlated electron materials: Dynamical mean‐field theory
  and electronic structure.
\newblock {\em AIP Conference Proceedings\/} {\bf 715} (2004) 3--74.
\newblock \doi{10.1063/1.1800733}.

\bibitem[{Kotliar et~al.(2006)Kotliar, Savrasov, Haule, Oudovenko, Parcollet,
  and Marianetti}]{Kotliar2006}
Kotliar G, Savrasov S, Haule K, Oudovenko V, Parcollet O, Marianetti C.
\newblock {Electronic structure calculations with dynamical mean-field theory}.
\newblock {\em Reviews of Modern Physics\/} {\bf 78} (2006) 865--951.
\newblock \doi{10.1103/RevModPhys.78.865}.

\bibitem[{Tamai et~al.(2019)Tamai, Zingl, Rozbicki, Cappelli, Ricc\`o, de~la
  Torre et~al.}]{Tamai:2019}
Tamai A, Zingl M, Rozbicki E, Cappelli E, Ricc\`o S, de~la Torre A, et~al.
\newblock High-resolution photoemission on
  ${\mathrm{sr}}_{2}{\mathrm{ruo}}_{4}$ reveals correlation-enhanced effective
  spin-orbit coupling and dominantly local self-energies.
\newblock {\em Phys. Rev. X\/} {\bf 9} (2019) 021048.
\newblock \doi{10.1103/PhysRevX.9.021048}.

\bibitem[{Karp et~al.(2020{\natexlab{c}})Karp, Bramberger, Grundner,
  Schollw\"ock, Millis, and Zingl}]{Karp:2020}
Karp J, Bramberger M, Grundner M, Schollw\"ock U, Millis AJ, Zingl M.
\newblock Sr$_2$moo$_4$ and sr$_2$ruo$_4$: Disentangling the roles of hund's
  and van hove physics.
\newblock {\em Phys. Rev. Lett.\/} {\bf 125} (2020{\natexlab{c}}) 166401.
\newblock \doi{10.1103/PhysRevLett.125.166401}.

\bibitem[{Perdew and Zunger(1981)}]{Perdew1981}
Perdew JP, Zunger A.
\newblock {Self-interaction correction to density-functional approximations for
  many-electron systems}.
\newblock {\em Physical Review B\/} {\bf 23} (1981) 5048--5079.
\newblock \doi{10.1103/PhysRevB.23.5048}.

\bibitem[{Burke and Wagner(2013)}]{Burke2013}
Burke K, Wagner LO.
\newblock Dft in a nutshell.
\newblock {\em International Journal of Quantum Chemistry\/} {\bf 113} (2013)
  96--101.
\newblock \doi{https://doi.org/10.1002/qua.24259}.

\bibitem[{Hampel et~al.(2020)Hampel, Beck, and
  Ederer}]{Hampel/Beck/Ederer:2020}
Hampel A, Beck S, Ederer C.
\newblock Effect of charge self-consistency in $\mathrm{DFT}+\mathrm{DMFT}$
  calculations for complex transition metal oxides.
\newblock {\em Phys. Rev. Research\/} {\bf 2} (2020) 033088.
\newblock \doi{10.1103/PhysRevResearch.2.033088}.

\bibitem[{Anisimov et~al.(2005)Anisimov, Kondakov, Kozhevnikov, Nekrasov,
  Pchelkina, Allen et~al.}]{anisimov2005full}
Anisimov VI, Kondakov DE, Kozhevnikov AV, Nekrasov IA, Pchelkina ZV, Allen JW,
  et~al.
\newblock Full orbital calculation scheme for materials with strongly
  correlated electrons.
\newblock {\em Phys. Rev. B\/} {\bf 71} (2005) 125119.
\newblock \doi{10.1103/PhysRevB.71.125119}.

\bibitem[{Amadon et~al.(2008)Amadon, Lechermann, Georges, Jollet, Wehling, and
  Lichtenstein}]{amadon2008plane}
Amadon B, Lechermann F, Georges A, Jollet F, Wehling TO, Lichtenstein AI.
\newblock Plane-wave based electronic structure calculations for correlated
  materials using dynamical mean-field theory and projected local orbitals.
\newblock {\em Phys. Rev. B\/} {\bf 77} (2008) 205112.
\newblock \doi{10.1103/PhysRevB.77.205112}.

\bibitem[{Aichhorn et~al.(2009)Aichhorn, Pourovskii, Vildosola, Ferrero,
  Parcollet, Miyake et~al.}]{aichhorn2009}
Aichhorn M, Pourovskii L, Vildosola V, Ferrero M, Parcollet O, Miyake T, et~al.
\newblock Dynamical mean-field theory within an augmented plane-wave framework:
  Assessing electronic correlations in the iron pnictide lafeaso.
\newblock {\em Phys. Rev. B\/} {\bf 80} (2009) 085101.
\newblock \doi{10.1103/PhysRevB.80.085101}.

\bibitem[{Marzari and Vanderbilt(1997)}]{Marzari/Vanderbilt:1997}
Marzari N, Vanderbilt D.
\newblock Maximally localized generalized wannier functions for composite
  energy bands.
\newblock {\em Phys. Rev. B\/} {\bf 56} (1997) 12847--12865.
\newblock \doi{10.1103/PhysRevB.56.12847}.

\bibitem[{Wang et~al.(2014)Wang, Lazar, Park, Millis, and Marianetti}]{Wang14}
Wang R, Lazar EA, Park H, Millis AJ, Marianetti CA.
\newblock {Selectively localized Wannier functions}.
\newblock {\em Phys. Rev. B\/} {\bf 90} (2014) 165125.
\newblock \doi{10.1103/PhysRevB.90.165125}.

\bibitem[{Marzari et~al.(2012)Marzari, Mostofi, Yates, Souza, and
  Vanderbilt}]{Marzari2012}
Marzari N, Mostofi AA, Yates JR, Souza I, Vanderbilt D.
\newblock {Maximally localized Wannier functions: Theory and applications}.
\newblock {\em Reviews of Modern Physics\/} {\bf 84} (2012) 1419--1475.
\newblock \doi{10.1103/RevModPhys.84.1419}.

\bibitem[{Souza et~al.(2001)Souza, Marzari, and
  Vanderbilt}]{Souza/Marzari/Vanderbilt:2001}
Souza I, Marzari N, Vanderbilt D.
\newblock Maximally localized wannier functions for entangled energy bands.
\newblock {\em Phys. Rev. B\/} {\bf 65} (2001) 035109.
\newblock \doi{10.1103/PhysRevB.65.035109}.

\bibitem[{Karp et~al.(2021{\natexlab{a}})Karp, Hampel, and
  Millis}]{Karp2021dependence}
Karp J, Hampel A, Millis AJ.
\newblock Dependence of $\mathrm{DFT}+\mathrm{DMFT}$ results on the
  construction of the correlated orbitals.
\newblock {\em Phys. Rev. B\/} {\bf 103} (2021{\natexlab{a}}) 195101.
\newblock \doi{10.1103/PhysRevB.103.195101}.

\bibitem[{Haule(2015)}]{Haule:2015}
Haule K.
\newblock Exact double counting in combining the dynamical mean field theory
  and the density functional theory.
\newblock {\em Phys. Rev. Lett.\/} {\bf 115} (2015) 196403.
\newblock \doi{10.1103/PhysRevLett.115.196403}.

\bibitem[{Beck et~al.(2021)Beck, Hampel, Parcollet, Ederer, and
  Georges}]{Beck:2021}
Beck S, Hampel A, Parcollet O, Ederer C, Georges A.
\newblock Charge self-consistent electronic structure calculations with
  dynamical mean-field theory using quantum espresso, wannier90 and triqs.
\newblock {\em arXiv\/}  (2021) 2111.10289.

\bibitem[{Park et~al.(2015)Park, Millis, and Marianetti}]{Park2015}
Park H, Millis AJ, Marianetti CA.
\newblock {Density functional versus spin-density functional and the choice of
  correlated subspace in multivariable effective action theories of electronic
  structure}.
\newblock {\em Phys. Rev. B\/} {\bf 92} (2015) 035146.

\bibitem[{Pavarini et~al.(2011)Pavarini, Koch, Lichtenstein, and
  Vollhardt}]{Pavarini:julich}
Pavarini E, Koch E, Lichtenstein A, Vollhardt DE.
\newblock {\em {T}he {LDA}+{DMFT} approach to strongly correlated materials\/},
  {\em Schriften des Forschungszentrums Jülich : Modeling and Simulation\/},
  vol.~1 (Schriften des Forschungszentrums Jülich) (2011).
\newblock Record converted from VDB: 12.11.2012.

\bibitem[{Slater(1960)}]{Slater1960}
Slater LJ.
\newblock {\em Confluent hypergeometric functions\/} (Cambridge Univ. Press)
  (1960).

\bibitem[{Vaugier et~al.(2012)Vaugier, Jiang, and Biermann}]{Vaugier2012}
Vaugier L, Jiang H, Biermann S.
\newblock Hubbard $u$ and hund exchange $j$ in transition metal oxides:
  Screening versus localization trends from constrained random phase
  approximation.
\newblock {\em Phys. Rev. B\/} {\bf 86} (2012) 165105.
\newblock \doi{10.1103/PhysRevB.86.165105}.

\bibitem[{Castellani et~al.(1978)Castellani, Natoli, and
  Ranninger}]{Castellani1978}
Castellani C, Natoli CR, Ranninger J.
\newblock Magnetic structure of v$_2$o$_3$ in the insulating phase.
\newblock {\em Phys. Rev. B\/} {\bf 18} (1978) 4945--4966.
\newblock \doi{10.1103/PhysRevB.18.4945}.

\bibitem[{Kanamori(1963)}]{Kanamori:1963}
Kanamori J.
\newblock {Electron Correlation and Ferromagnetism of Transition Metals}.
\newblock {\em Progress of Theoretical Physics\/} {\bf 30} (1963) 275--289.
\newblock \doi{10.1143/PTP.30.275}.

\bibitem[{Anisimov and Gunnarsson(1991)}]{Anisimov:1991}
Anisimov VI, Gunnarsson O.
\newblock Density-functional calculation of effective coulomb interactions in
  metals.
\newblock {\em Phys. Rev. B\/} {\bf 43} (1991) 7570--7574.
\newblock \doi{10.1103/PhysRevB.43.7570}.

\bibitem[{Aryasetiawan et~al.(2004)Aryasetiawan, Imada, Georges, Kotliar,
  Biermann, and Lichtenstein}]{PhysRevB.70.195104}
Aryasetiawan F, Imada M, Georges A, Kotliar G, Biermann S, Lichtenstein AI.
\newblock Frequency-dependent local interactions and low-energy effective
  models from electronic structure calculations.
\newblock {\em Phys. Rev. B\/} {\bf 70} (2004) 195104.
\newblock \doi{10.1103/PhysRevB.70.195104}.

\bibitem[{Aryasetiawan et~al.(2011)Aryasetiawan, Miyake, and
  Sakuma}]{Aryasetiawan:2011}
Aryasetiawan F, Miyake T, Sakuma R.
\newblock {\em {T}he {LDA}+{DMFT} approach to strongly correlated materials\/}
  (Schriften des Forschungszentrums Jülich), chap. The Constrained RPA Method
  for Calculating the Hubard $U$ from First-Principles (2011), pp. 7.1 -- 7.26.

\bibitem[{Kaltak(2015)}]{Merzuk2015}
Kaltak M.
\newblock {\em Merging GW with DMFT\/}.
\newblock Ph.D. thesis, University of Vienna (2015).

\bibitem[{Vogel et~al.(1996)Vogel, Kr\"uger, and Pollmann}]{Vogel1996}
Vogel D, Kr\"uger P, Pollmann J.
\newblock {Self-interaction and relaxation-corrected pseudopotentials for II-VI
  semiconductors}.
\newblock {\em Phys. Rev. B\/} {\bf 54} (1996) 5495--5511.

\bibitem[{Filippetti and Spaldin(2003)}]{Filippetti2003}
Filippetti A, Spaldin NA.
\newblock Self-interaction-corrected pseudopotential scheme for magnetic and
  strongly-correlated systems.
\newblock {\em Phys. Rev. B\/} {\bf 67} (2003) 125109.
\newblock \doi{10.1103/PhysRevB.67.125109}.

\bibitem[{K\"orner and Els\"asser(2010)}]{Koerner2010}
K\"orner W, Els\"asser C.
\newblock {First-principles density functional study of dopant elements at
  grain boundaries in ZnO}.
\newblock {\em Phys. Rev. B\/} {\bf 81} (2010) 085324.

\bibitem[{Lechermann et~al.(2019)Lechermann, K\"orner, Urban, and
  Els\"asser}]{Lechermann2019}
Lechermann F, K\"orner W, Urban DF, Els\"asser C.
\newblock Interplay of charge-transfer and mott-hubbard physics approached by
  an efficient combination of self-interaction correction and dynamical
  mean-field theory.
\newblock {\em Phys. Rev. B\/} {\bf 100} (2019) 115125.

\bibitem[{Wang et~al.(2020{\natexlab{b}})Wang, Zheng, Krivyakina, Chmaissem,
  Lopes, Lynn et~al.}]{Wang2020}
Wang BX, Zheng H, Krivyakina E, Chmaissem O, Lopes PP, Lynn JW, et~al.
\newblock {Synthesis and characterization of bulk Nd$_{1-x}$Sr$_x$NiO$_2$ and
  Nd$_{1-x}$Sr$_x$NiO$_3$}.
\newblock {\em Physical Review Materials\/} {\bf 4} (2020{\natexlab{b}})
  084409.
\newblock \doi{10.1103/PhysRevMaterials.4.084409}.

\bibitem[{Si et~al.(2020)Si, Xiao, Kaufmann, Tomczak, Lu, Zhong
  et~al.}]{Si2020}
Si L, Xiao W, Kaufmann J, Tomczak JM, Lu Y, Zhong Z, et~al.
\newblock {Topotactic Hydrogen in Nickelate Superconductors and Akin
  Infinite-Layer Oxides $AB$O$_2$}.
\newblock {\em Physical Review Letters\/} {\bf 124} (2020) 166402.
\newblock \doi{10.1103/PhysRevLett.124.166402}.

\bibitem[{Geisler and Pentcheva(2020)}]{Geisler2020}
Geisler B, Pentcheva R.
\newblock {Fundamental difference in the electronic reconstruction of
  infinite-layer versus perovskite neodymium nickelate films on SrTiO$_3$
  (001)}.
\newblock {\em Physical Review B\/} {\bf 102} (2020) 020502.
\newblock \doi{10.1103/PhysRevB.102.020502}.

\bibitem[{Sakakibara et~al.(2020)Sakakibara, Usui, Suzuki, Kotani, Aoki, and
  Kuroki}]{Sakakibara2020}
Sakakibara H, Usui H, Suzuki K, Kotani T, Aoki H, Kuroki K.
\newblock {Model Construction and a Possibility of Cupratelike Pairing in a New
  $d^9$ Nickelate Superconductor (Nd, Sr)NiO$_2$}.
\newblock {\em Physical Review Letters\/} {\bf 125} (2020) 077003.
\newblock \doi{10.1103/PhysRevLett.125.077003}.

\bibitem[{He et~al.(2020)He, Jiang, Lu, Song, Chen, Jin et~al.}]{He2020}
He R, Jiang P, Lu Y, Song Y, Chen M, Jin M, et~al.
\newblock {Polarity-induced electronic and atomic reconstruction at
  NdNiO$_2$/SrTiO$_3$ interfaces}.
\newblock {\em Physical Review B\/} {\bf 102} (2020) 035118.
\newblock \doi{10.1103/PhysRevB.102.035118}.

\bibitem[{Zhang et~al.(2020{\natexlab{a}})Zhang, Jin, Wang, Xi, Shi, Ye
  et~al.}]{Zhang2020}
Zhang H, Jin L, Wang S, Xi B, Shi X, Ye F, et~al.
\newblock {Effective Hamiltonian for nickelate oxides Nd$_{1-x}$Sr$_x$NiO$_2$}.
\newblock {\em Physical Review Research\/} {\bf 2} (2020{\natexlab{a}}) 013214.
\newblock \doi{10.1103/PhysRevResearch.2.013214}.

\bibitem[{Zhang et~al.(2020{\natexlab{b}})Zhang, Zheng, Chen, Ren, Yonemura,
  Huq et~al.}]{Zhang2020c}
Zhang J, Zheng H, Chen YS, Ren Y, Yonemura M, Huq A, et~al.
\newblock {High oxygen pressure floating zone growth and crystal structure of
  the metallic nickelates $R_4$Ni$_3$O$_{10}$ ($R$ = La , Pr)}.
\newblock {\em Physical Review Materials\/} {\bf 4} (2020{\natexlab{b}})
  083402.
\newblock \doi{10.1103/PhysRevMaterials.4.083402}.

\bibitem[{Bernardini et~al.(2020)Bernardini, Olevano, Blase, and
  Cano}]{Bernardini2020}
Bernardini F, Olevano V, Blase X, Cano A.
\newblock {Infinite-layer fluoro-nickelates as $d^9$ model materials}.
\newblock {\em Journal of Physics: Materials\/} {\bf 3} (2020) 035003.
\newblock \doi{10.1088/2515-7639/ab885d}.

\bibitem[{Bernardini and Cano(2020)}]{Bernardini2020a}
Bernardini F, Cano A.
\newblock {Stability and electronic properties of LaNiO$_2$/SrTiO$_3$
  heterostructures}.
\newblock {\em Journal of Physics: Materials\/} {\bf 3} (2020) 03LT01.
\newblock \doi{10.1088/2515-7639/ab9d0f}.

\bibitem[{Liu et~al.(2021)Liu, Xu, Cao, Zhu, Wang, and Yang}]{Liu2021}
Liu Z, Xu C, Cao C, Zhu W, Wang ZF, Yang J.
\newblock {Doping dependence of electronic structure of infinite-layer
  NdNiO$_2$}.
\newblock {\em Physical Review B\/} {\bf 103} (2021) 045103.
\newblock \doi{10.1103/PhysRevB.103.045103}.

\bibitem[{Plienbumrung et~al.(2021)Plienbumrung, Daghofer, and
  Ole{\'{s}}}]{Plienbumrung2021}
Plienbumrung T, Daghofer M, Ole{\'{s}} AM.
\newblock {Interplay between Zhang-Rice singlets and high-spin states in a
  model for doped NiO$_2$ planes}.
\newblock {\em Physical Review B\/} {\bf 103} (2021) 104513.
\newblock \doi{10.1103/PhysRevB.103.104513}.

\bibitem[{Malyi et~al.(2021)Malyi, Varignon, and Zunger}]{Malyi2021}
Malyi OI, Varignon J, Zunger A.
\newblock {Bulk NdNiO2 is thermodynamically unstable with respect to
  decomposition while hydrogenation reduces the instability and transforms it
  from metal to insulator}.
\newblock {\em arXiv\/}  (2021) 2107.01790.

\bibitem[{Peng et~al.(2021)Peng, Jiang, Moritz, Devereaux, and Jia}]{Peng2021}
Peng C, Jiang HC, Moritz B, Devereaux TP, Jia C.
\newblock {Superconductivity in a minimal two-band model for infinite-layer
  nickelates}.
\newblock {\em arXiv\/}  (2021) 2110.07593.

\bibitem[{Choubey and Eremin(2021)}]{Choubey2021}
Choubey P, Eremin IM.
\newblock {Electronic theory for scanning tunneling microscopy spectra in
  infinite-layer nickelate superconductors}.
\newblock {\em Physical Review B\/} {\bf 104} (2021) 144504.
\newblock \doi{10.1103/PhysRevB.104.144504}.

\bibitem[{Sawatzky(2019)}]{Sawatzky2019}
Sawatzky GA.
\newblock {Superconductivity seen in a non-magnetic nickel oxide}.
\newblock {\em Nature\/} {\bf 572} (2019).
\newblock \doi{10.1038/d41586-019-02518-3}.

\bibitem[{Nomura et~al.(2020)Nomura, Nomoto, Hirayama, and Arita}]{Nomura2020}
Nomura Y, Nomoto T, Hirayama M, Arita R.
\newblock {Magnetic exchange coupling in cuprate-analog $d^9$ nickelates}.
\newblock {\em Physical Review Research\/} {\bf 2} (2020) 043144.
\newblock \doi{10.1103/PhysRevResearch.2.043144}.

\bibitem[{Hirayama et~al.(2020)Hirayama, Tadano, Nomura, and
  Arita}]{Hirayama2020}
Hirayama M, Tadano T, Nomura Y, Arita R.
\newblock {Materials design of dynamically stable $d^9$ layered nickelates}.
\newblock {\em Physical Review B\/} {\bf 101} (2020) 075107.
\newblock \doi{10.1103/PhysRevB.101.075107}.

\bibitem[{Jin et~al.(2020)Jin, Pickett, and Lee}]{Jin2020}
Jin HS, Pickett WE, Lee KW.
\newblock {Proposed ordering of textured spin singlets in a bulk infinite-layer
  nickelate}.
\newblock {\em Physical Review Research\/} {\bf 2} (2020) 033197.
\newblock \doi{10.1103/PhysRevResearch.2.033197}.

\bibitem[{Petocchi et~al.(2020)Petocchi, Christiansson, Nilsson, Aryasetiawan,
  and Werner}]{Petocchi2020a}
Petocchi F, Christiansson V, Nilsson F, Aryasetiawan F, Werner P.
\newblock {Normal State of Nd$_{1-x}$Sr$_x$NiO$_2$ from Self-Consistent $GW$ +
  EDMFT}.
\newblock {\em Physical Review X\/} {\bf 10} (2020) 041047.
\newblock \doi{10.1103/PhysRevX.10.041047}.

\bibitem[{Higashi et~al.(2021)Higashi, Winder, Kune{\v{s}}, and
  Hariki}]{Higashi2021}
Higashi K, Winder M, Kune{\v{s}} J, Hariki A.
\newblock {Core-Level X-Ray Spectroscopy of Infinite-Layer Nickelate: LDA +
  DMFT Study}.
\newblock {\em Physical Review X\/} {\bf 11} (2021) 041009.
\newblock \doi{10.1103/PhysRevX.11.041009}.

\bibitem[{Leonov(2021)}]{Leonov2021}
Leonov I.
\newblock {Effect of lattice strain on the electronic structure and magnetic
  correlations in infinite-layer (Nd,Sr)NiO$_2$}.
\newblock {\em Journal of Alloys and Compounds\/} {\bf 883} (2021) 160888.
\newblock \doi{10.1016/J.JALLCOM.2021.160888}.

\bibitem[{Lin et~al.(2021)Lin, Gawryluk, Klein, Huangfu, Pomjakushina, von Rohr
  et~al.}]{Lin2021a}
Lin H, Gawryluk DJ, Klein YM, Huangfu S, Pomjakushina E, von Rohr F, et~al.
\newblock {Universal spin-glass behaviour in bulk LaNiO$_2$, PrNiO$_2$ and
  NdNiO$_2$}.
\newblock {\em arXiv\/}  (2021) 2104.14324.

\bibitem[{Lee and Pickett(2004)}]{Lee2004}
Lee KW, Pickett WE.
\newblock {Infinite-layer LaNiO$_2$: Ni$^{1+}$ is not Cu$^{2+}$}.
\newblock {\em Physical Review B\/} {\bf 70} (2004) 165109.
\newblock \doi{10.1103/PhysRevB.70.165109}.

\bibitem[{Choi et~al.(2020{\natexlab{a}})Choi, Pickett, and Lee}]{Choi2020a}
Choi MY, Pickett WE, Lee KW.
\newblock {Fluctuation-frustrated flat band instabilities in NdNiO$_2$}.
\newblock {\em Physical Review Research\/} {\bf 2} (2020{\natexlab{a}}) 033445.
\newblock \doi{10.1103/PhysRevResearch.2.033445}.

\bibitem[{Choi et~al.(2020{\natexlab{b}})Choi, Lee, and Pickett}]{Choi2020}
Choi MY, Lee KW, Pickett WE.
\newblock {Role of 4$f$ states in infinite-layer NdNiO$_2$}.
\newblock {\em Physical Review B\/} {\bf 101} (2020{\natexlab{b}}) 020503.
\newblock \doi{10.1103/PhysRevB.101.020503}.

\bibitem[{Karp et~al.(2021{\natexlab{b}})Karp, Hampel, and Millis}]{Karp:2021}
Karp J, Hampel A, Millis AJ.
\newblock Dependence of $\mathrm{DFT}+\mathrm{DMFT}$ results on the
  construction of the correlated orbitals.
\newblock {\em Phys. Rev. B\/} {\bf 103} (2021{\natexlab{b}}) 195101.
\newblock \doi{10.1103/PhysRevB.103.195101}.

\bibitem[{Zeng et~al.(2020)Zeng, Tang, Yin, Li, Li, Huang et~al.}]{Zeng2020}
Zeng S, Tang CS, Yin X, Li C, Li M, Huang Z, et~al.
\newblock {Phase Diagram and Superconducting Dome of Infinite-Layer
  Nd$_{1-x}$Sr$_x$NiO$_2$ Thin Films}.
\newblock {\em Physical Review Letters\/} {\bf 125} (2020) 147003.
\newblock \doi{10.1103/PhysRevLett.125.147003}.

\bibitem[{Li et~al.(2020)Li, Wang, Lee, Harvey, Osada, Goodge et~al.}]{Li2020a}
Li D, Wang BY, Lee K, Harvey SP, Osada M, Goodge BH, et~al.
\newblock {Superconducting Dome in Nd$_{1-x}$Sr$_x$NiO$_2$ Infinite Layer
  Films}.
\newblock {\em Physical Review Letters\/} {\bf 125} (2020) 027001.
\newblock \doi{10.1103/PhysRevLett.125.027001}.

\bibitem[{Gu et~al.(2020{\natexlab{b}})Gu, Li, Wan, Li, Guo, Yang
  et~al.}]{Gu2020a}
Gu Q, Li Y, Wan S, Li H, Guo W, Yang H, et~al.
\newblock {Single particle tunneling spectrum of superconducting
  Nd$_{1-x}$Sr$_x$NiO$_2$ thin films}.
\newblock {\em Nature Communications\/} {\bf 11} (2020{\natexlab{b}}).
\newblock \doi{10.1038/s41467-020-19908-1}.

\bibitem[{Goodge et~al.(2021)Goodge, Li, Lee, Osada, Wang, Sawatzky
  et~al.}]{Goodge2021}
Goodge BH, Li D, Lee K, Osada M, Wang BY, Sawatzky GA, et~al.
\newblock {Doping evolution of the Mott-Hubbard landscape in infinite-layer
  nickelates.}
\newblock {\em Proceedings of the National Academy of Sciences of the United
  States of America\/} {\bf 118} (2021).
\newblock \doi{10.1073/pnas.2007683118}.

\bibitem[{Wang et~al.(2021)Wang, Li, Goodge, Lee, Osada, Harvey
  et~al.}]{Wang2021}
Wang BY, Li D, Goodge BH, Lee K, Osada M, Harvey SP, et~al.
\newblock {Isotropic Pauli-limited superconductivity in the infinite-layer
  nickelate Nd$_{0.775}$Sr$_{0.225}$NiO$_2$}.
\newblock {\em Nature Physics\/} {\bf 17} (2021).
\newblock \doi{10.1038/s41567-020-01128-5}.

\bibitem[{Zhao et~al.(2021)Zhao, Zhou, Fu, Wang, Zhou, Cheng et~al.}]{Zhao2021}
Zhao D, Zhou YB, Fu Y, Wang L, Zhou XF, Cheng H, et~al.
\newblock {Intrinsic Spin Susceptibility and Pseudogaplike Behavior in
  Infinite-Layer LaNiO$_2$}.
\newblock {\em Physical Review Letters\/} {\bf 126} (2021) 197001.
\newblock \doi{10.1103/PhysRevLett.126.197001}.

\bibitem[{Lu et~al.(2021)Lu, Rossi, Nag, Osada, Li, Lee et~al.}]{Lu2021}
Lu H, Rossi M, Nag A, Osada M, Li DF, Lee K, et~al.
\newblock {Magnetic excitations in infinite-layer nickelates}.
\newblock {\em Science\/} {\bf 373} (2021).
\newblock \doi{10.1126/science.abd7726}.

\bibitem[{Osada et~al.(2020{\natexlab{a}})Osada, {Yang Wang}, {H. Goodge}, Lee,
  Yoon, Sakuma et~al.}]{Osada2020}
Osada M, {Yang Wang} B, {H Goodge} B, Lee K, Yoon H, Sakuma K, et~al.
\newblock {A Superconducting Praseodymium Nickelate with Infinite Layer
  Structure}.
\newblock {\em Nano Letters\/} {\bf 20} (2020{\natexlab{a}}) 5735--5740.
\newblock \doi{10.1021/acs.nanolett.0c01392}.

\bibitem[{Osada et~al.(2020{\natexlab{b}})Osada, Wang, Lee, Li, and
  Hwang}]{Osada2020a}
Osada M, Wang BY, Lee K, Li D, Hwang HY.
\newblock {Phase diagram of infinite layer praseodymium nickelate
  Pr$_{1-x}$Sr$_x$NiO$_2$ thin films}.
\newblock {\em Physical Review Materials\/} {\bf 4} (2020{\natexlab{b}})
  121801.
\newblock \doi{10.1103/PhysRevMaterials.4.121801}.

\bibitem[{Osada et~al.(2021)Osada, Wang, Goodge, Harvey, Lee, Li
  et~al.}]{Osada2021}
Osada M, Wang BY, Goodge BH, Harvey SP, Lee K, Li D, et~al.
\newblock {Nickelate Superconductivity without Rare‐Earth Magnetism:
  (La,Sr)NiO$_2$}.
\newblock {\em Advanced Materials\/} {\bf 33} (2021) 2104083.
\newblock \doi{10.1002/adma.202104083}.

\bibitem[{Ren et~al.(2021)Ren, Gao, Zhao, Luo, Zhou, and Zhu}]{Ren2021}
Ren X, Gao Q, Zhao Y, Luo H, Zhou X, Zhu Z.
\newblock {Superconductivity in infinite-layer Pr$_{0.8}$Sr$_{0.2}$NiO$_2$
  films on different substrates}.
\newblock {\em arXiv\/}  (2021) 2109.05761.

\bibitem[{Zeng et~al.(2021)Zeng, Li, Chow, Cao, Zhang, Tang et~al.}]{Zeng2021}
Zeng SW, Li CJ, Chow LE, Cao Y, Zhang ZT, Tang CS, et~al.
\newblock {Superconductivity in infinite-layer lanthanide nickelates}.
\newblock {\em arXiv\/}  (2021) 2105.13492.

\bibitem[{Puphal et~al.(2021)Puphal, Wu, F{\"{u}}rsich, Lee, Pakdaman, Bruin
  et~al.}]{Puphal2021}
Puphal P, Wu YM, F{\"{u}}rsich K, Lee H, Pakdaman M, Bruin JAN, et~al.
\newblock {Synthesis and Characterization of Ca-Substituted Infinite-Layer
  Nickelate Crystals}.
\newblock {\em arXiv\/}  (2021) 2106.13171.

\bibitem[{Liu et~al.(2020)Liu, Ren, Zhu, Wang, and Yang}]{Liu2020}
Liu Z, Ren Z, Zhu W, Wang Z, Yang J.
\newblock {Electronic and magnetic structure of infinite-layer NdNiO$_2$: trace
  of antiferromagnetic metal}.
\newblock {\em npj Quantum Materials\/} {\bf 5} (2020).
\newblock \doi{10.1038/s41535-020-0229-1}.

\bibitem[{Jiang et~al.(2019)Jiang, Si, Liao, and Zhong}]{Jiang2019}
Jiang P, Si L, Liao Z, Zhong Z.
\newblock {Electronic structure of rare-earth infinite-layer $R$NiO$_2$ ($R$=
  La , Nd )}.
\newblock {\em Physical Review B\/} {\bf 100} (2019) 201106.
\newblock \doi{10.1103/PhysRevB.100.201106}.

\bibitem[{Ryee et~al.(2020)Ryee, Yoon, Kim, Jeong, and Han}]{Ryee2020}
Ryee S, Yoon H, Kim TJ, Jeong MY, Han MJ.
\newblock {Induced magnetic two-dimensionality by hole doping in the
  superconducting infinite-layer nickelate Nd$_{1-x}$Sr$_x$NiO$_2$}.
\newblock {\em Physical Review B\/} {\bf 101} (2020) 064513.
\newblock \doi{10.1103/PhysRevB.101.064513}.

\bibitem[{Zhang et~al.(2020{\natexlab{c}})Zhang, Yang, and Zhang}]{Zhang2020a}
Zhang GM, Yang Yf, Zhang FC.
\newblock {Self-doped Mott insulator for parent compounds of nickelate
  superconductors}.
\newblock {\em Physical Review B\/} {\bf 101} (2020{\natexlab{c}}) 020501.
\newblock \doi{10.1103/PhysRevB.101.020501}.

\bibitem[{Yang and Zhang(2021)}]{Yang2021}
Yang Yf, Zhang GM.
\newblock {Self-doping and the Mott-Kondo scenario for infinite-layer nickelate
  superconductors}.
\newblock {\em arXiv\/}  (2021) 2110.11089.

\bibitem[{Xia et~al.(2021)Xia, Wu, Chen, and Chen}]{Xia2021}
Xia C, Wu J, Chen Y, Chen H.
\newblock {Dynamical structural instability and a new crystal-electronic
  structure of infinite-layer nickelates}.
\newblock {\em arXiv\/}  (2021) 2110.12405.

\bibitem[{Zaanen et~al.(1985)Zaanen, Sawatzky, and Allen}]{Zaanen1985}
Zaanen J, Sawatzky GA, Allen JW.
\newblock {Band gaps and electronic structure of transition-metal compounds}.
\newblock {\em Physical Review Letters\/} {\bf 55} (1985) 418--421.

\bibitem[{Lechermann(2020{\natexlab{b}})}]{Lechermann2020}
Lechermann F.
\newblock {Late transition metal oxides with infinite-layer structure:
  Nickelates versus cuprates}.
\newblock {\em Physical Review B\/} {\bf 101} (2020{\natexlab{b}}) 081110.
\newblock \doi{10.1103/PhysRevB.101.081110}.

\bibitem[{Shen et~al.(2021)Shen, Sears, Fabbris, Li, Pelliciari, Jarrige
  et~al.}]{Shen2021}
Shen Y, Sears J, Fabbris G, Li J, Pelliciari J, Jarrige I, et~al.
\newblock {Role of oxygen states in square planar $d^{9-\delta}$ nickelates}
  (2021).

\bibitem[{Lechermann(2021{\natexlab{a}})}]{Lechermann2021}
Lechermann F.
\newblock {Doping-dependent character and possible magnetic ordering of
  NdNiO$_2$}.
\newblock {\em Phys. Rev. Materials\/} {\bf 5} (2021{\natexlab{a}}) 044803.

\bibitem[{Stadler et~al.(2019)Stadler, Kotliar, Weichselbaum, and {von
  Delft}}]{Stadler19}
Stadler K, Kotliar G, Weichselbaum A, {von Delft} J.
\newblock Hundness versus mottness in a three-band hubbard–hund model: On the
  origin of strong correlations in hund metals.
\newblock {\em Annals of Physics\/} {\bf 405} (2019) 365--409.
\newblock \doi{https://doi.org/10.1016/j.aop.2018.10.017}.

\bibitem[{Karp et~al.(2020{\natexlab{d}})Karp, Bramberger, Grundner,
  Schollw\"ock, Millis, and Zingl}]{Karp2020b}
Karp J, Bramberger M, Grundner M, Schollw\"ock U, Millis AJ, Zingl M.
\newblock Sr$_2$moo$_4$ and sr$_2$ruo$_4$: Disentangling the roles of hund's
  and van hove physics.
\newblock {\em Phys. Rev. Lett.\/} {\bf 125} (2020{\natexlab{d}}) 166401.
\newblock \doi{10.1103/PhysRevLett.125.166401}.

\bibitem[{Kang et~al.(2020)Kang, Melnick, Semon, Kotliar, and
  Choi}]{kang2020infinitelayer}
[Dataset] Kang B, Melnick C, Semon P, Kotliar G, Choi S (2020).

\bibitem[{Zhang et~al.(2021)Zhang, Lane, Singh, Nokelainen, Barbiellini,
  Markiewicz et~al.}]{Zhang2021}
Zhang R, Lane C, Singh B, Nokelainen J, Barbiellini B, Markiewicz RS, et~al.
\newblock {Magnetic and $f$-electron effects in LaNiO$_2$ and NdNiO$_2$
  nickelates with cuprate-like $3d_{x^2-y^2}$ band}.
\newblock {\em Communications Physics\/} {\bf 4} (2021).
\newblock \doi{10.1038/s42005-021-00621-4}.

\bibitem[{Katukuri et~al.(2020)Katukuri, Bogdanov, Weser, van~den Brink, and
  Alavi}]{Katukuri2020}
Katukuri VM, Bogdanov NA, Weser O, van~den Brink J, Alavi A.
\newblock {Electronic correlations and magnetic interactions in infinite-layer
  NdNiO$_2$}.
\newblock {\em Physical Review B\/} {\bf 102} (2020) 241112.
\newblock \doi{10.1103/PhysRevB.102.241112}.

\bibitem[{Wang et~al.(2020{\natexlab{c}})Wang, Zhang, Yang, and
  Zhang}]{Wang2020a}
Wang Z, Zhang GM, Yang Yf, Zhang FC.
\newblock {Distinct pairing symmetries of superconductivity in infinite-layer
  nickelates}.
\newblock {\em Physical Review B\/} {\bf 102} (2020{\natexlab{c}}) 220501.
\newblock \doi{10.1103/PhysRevB.102.220501}.

\bibitem[{Adhikary et~al.(2020)Adhikary, Bandyopadhyay, Das, Dasgupta, and
  Saha-Dasgupta}]{Adhikary2020}
Adhikary P, Bandyopadhyay S, Das T, Dasgupta I, Saha-Dasgupta T.
\newblock {Orbital-selective superconductivity in a two-band model of
  infinite-layer nickelates}.
\newblock {\em Physical Review B\/} {\bf 102} (2020) 100501.
\newblock \doi{10.1103/PhysRevB.102.100501}.

\bibitem[{Momma and Izumi(2011)}]{vesta}
Momma K, Izumi F.
\newblock {{\it VESTA3} for three-dimensional visualization of crystal,
  volumetric and morphology data}.
\newblock {\em Journal of Applied Crystallography\/} {\bf 44} (2011)
  1272--1276.
\newblock \doi{10.1107/S0021889811038970}.

\bibitem[{Lechermann(2021{\natexlab{b}})}]{Lechermann2021a}
Lechermann F.
\newblock {\em npj Computational Materials\/} {\bf 7} (2021{\natexlab{b}}) 120.

\bibitem[{Leonov et~al.(2020)Leonov, Skornyakov, and Savrasov}]{Leonov2020}
Leonov I, Skornyakov SL, Savrasov SY.
\newblock {Lifshitz transition and frustration of magnetic moments in
  infinite-layer NdNiO 2 upon hole doping}.
\newblock {\em Physical Review B\/} {\bf 101} (2020) 241108.
\newblock \doi{10.1103/PhysRevB.101.241108}.

\end{thebibliography}


\section*{Figures}

\begin{table}
\centering
\begin{tabular}{c c c c c c c c}
\hline\hline
             & $d_{x^2-y^2}$ & $d_{3z^2-r^2}$ && LS $N$=2 & HS $N$=2 & $N$=3  & $N$=4  \\ \hline
MLWF           & 1.13          & 1.91      && 0.04   & 0.05   & 0.78 & 0.13 \\ 
SLWF           & 1.27          & 1.93      && 0.03   & 0.02   & 0.69 & 0.26 \\ 
Proj -10 to 10 & 1.14          & 1.65      && 0.11   & 0.15   & 0.64 & 0.09 \\ 
Proj -10 to 3  & 1.15          & 1.81      && 0.07   & 0.08   & 0.72 & 0.12 \\ \hline\hline
\end{tabular}
\caption{Orbital occupancies  of the most relevant Ni-$d$ orbitals of NdNiO$_2$ from the Matsubara Green function  (left) and occurrence probabilities of low spin (LS $S = 0$) and high spin (HS $S = 1$) multiplet configurations obtained from the impurity density matrix computed for stoichiometric NdNiO$_2$ using a Kanamori Hamiltonian with two correlated orbitals and $\mathcal{U} = 7$ eV and $\mathcal{J} = 0.7$ eV at $T = 290$ K.  From Ref.~\cite{Karp2021dependence}. }
\label{tab:dmft_occ}
\end{table}

\begin{table}
\centering
\begin{tabular}{l l l l l l}
\hline\hline
Ref.                  & downfolding model         & $n_c$ & interactions (eV) & $T$ (K) & $d_{x^2-y^2}$ $m^*/m$ \\ \hline
\cite{Karp2020}       & 3 MLWF                    & 1     & $\mathcal{U}$ = 3.1                       & 290   & 4.0         \\ 
\cite{Gu2020}         & 4 MLWF                    & 1     & $\mathcal{U}$ = 3.0                       & 116   & 3.3         \\
\cite{Petocchi2020a}  & 7 MLWF (5 Ni + 2 Nd)      & 5     & full GW+EDMFT                             & 1160  &  5.6         \\
\cite{Kitatani2020}   & 10 MLWF (5 Ni + 5 Nd)     & 2     & $\mathcal{U}$ = 3.1, $\mathcal{J}$ = 0.65 & 300   & 4.4         \\ 
\cite{Karp:2021}      & 13 MLWF                   & 2     & $\mathcal{U}$ = 7, $\mathcal{J}$ = 0.7    & 290   & 7.6         \\ 
\cite{Karp:2021}      & 13 SLWF                   & 2     & $\mathcal{U}$ =7, $\mathcal{J}$ = 0.7     & 290   & 3.9         \\ 
\cite{Leonov2020}     & 16 MLWF (Ni, Nd, O)      & 5     & $U_{avg}$ = 6, $J_{avg}$ = 0.95           & 290   & 3           \\ 
\cite{Karp:2021}      & projectors -10 to 3       & 2     & $\mathcal{U}$ = 7, $\mathcal{J}$ = 0.7    & 290   & 5.6         \\ 
\cite{Lechermann2021} & projectors -10 to 3       & 5     & $U_{avg}$ = 10, $J_{avg}$ = 1             & 30    & 6.4        \\ 
\cite{Lechermann2021} & projectors -10 to 3       & 5     & $U_{avg}$ = 10, $J_{avg}$ = 1  (SIC on O) & 30    & Mott insulating         \\ 
\cite{Karp:2021}      & projectors -10 to 10      & 2     & $\mathcal{U}$ = 7, $\mathcal{J}$ = 0.7    & 290   & 4.6         \\ 
\cite{Karp2020a}      & projectors -10 to 10      & 5     & $U_{avg}$ = 7, $J_{avg}$ = 0.7            & 390   & 3.7         \\ 
\cite{Wang2020b}      & projectors -10 to 10      & 5     & $U_{avg}$ = 5, $J_{avg}$ = 1              & 100   & 2.8         \\ 
\cite{Ryee2020}       & projectors -10 to 10      & 5     & $U_{avg}$ = 5, $J_{avg}$ = 0.8            & -     & 2.4         \\ 
\cite{Ryee2020}       & projectors -10 to 10      & 5     & $U_{avg}$ = 9, $J_{avg}$ = 0.8            & -     & 4.1         \\ 
\cite{Kang2021}      & projectors -10 to 10      & 5     & $U_{avg}$ = 5, $J_{avg}$ = 1              & 60   & 2.6         \\ 

\hline\hline

\end{tabular}
\caption{$d_{x^2-y^2}$ orbital mass enhancement from different DMFT results in the literature. $n_c$ refers to the number of correlated orbitals in the impurity problem.}
\label{tab:mass_enhancements}
\end{table}

\end{document}